\documentclass{llncs}
\usepackage{graphicx}
\usepackage{xcolor}
\usepackage{listings}
\usepackage{amsfonts}
\usepackage{amsmath}

\usepackage{algorithm,algorithmic}
\usepackage[colorlinks=true,hyperfootnotes=false]{hyperref}
\usepackage[capitalise]{cleveref}
\usepackage{siunitx}
\usepackage{todonotes}
\usepackage{cite}
\usepackage[
	n,
	operators,
	probability]{cryptocode}
\usepackage{ifthen}
\newboolean{submission}
\setboolean{submission}{false}
\usepackage{comment}
\usepackage{booktabs}
\usepackage{multirow}
\usepackage{qcircuit}
\usepackage{etoolbox}
\usepackage{tabularray}
\UseTblrLibrary{booktabs}

\usepackage{comment}
\newtoggle{comments}
\toggletrue{comments}
\iftoggle{comments}{
	\newcommand{\rv}[2][]{\todo[color=green!30,#1]{\textsf{RV:} #2}}
        
}{
    \newcommand{\rv}[2]{}
}
\lstdefinestyle{mystyle}{
    backgroundcolor=\color{lightgray},   
    basicstyle=\ttfamily\footnotesize,   
    keywordstyle=\bfseries\color{blue},  
    commentstyle=\color{green},          
    stringstyle=\color{red},             
    numberstyle=\tiny\color{gray},       
    breaklines=true,                     
    frame=single,                        
    rulecolor=\color{black},             
    numbers=left,                        
    numbersep=5pt,                       
    showstringspaces=false               
}

\newcommand{\FF}{\mathbb{F}}

\newcommand{\cP}{\mathcal{P}}

\definecolor{codegreen}{rgb}{0,0.6,0}
\definecolor{codegray}{rgb}{0.5,0.5,0.5}
\definecolor{codepurple}{rgb}{0.58,0,0.82}
\definecolor{backcolour}{rgb}{1.0,1.0,1.0}

\lstdefinestyle{mystyle}{
    backgroundcolor=\color{backcolour},   
    commentstyle=\color{codegreen},
    keywordstyle=\color{magenta},
    numberstyle=\tiny\color{codegray},
    stringstyle=\color{codepurple},
    basicstyle=\ttfamily\footnotesize,
    breakatwhitespace=false,         
    breaklines=true,                 
    captionpos=b,                    
    keepspaces=true,                 
    numbers=left,                    
    numbersep=5pt,                  
    showspaces=false,                
    showstringspaces=false,
    showtabs=false,                  
    tabsize=2
}

\mathchardef\mhyphen="2D                

\title{ A Fault Analysis on SNOVA}
\ifthenelse{\boolean{submission}}
{}{
\author{Gustavo Banegas$^{1}\href{https://orcid.org/0000-0001-5502-2977} {\includegraphics[height=\fontcharht\font`B]{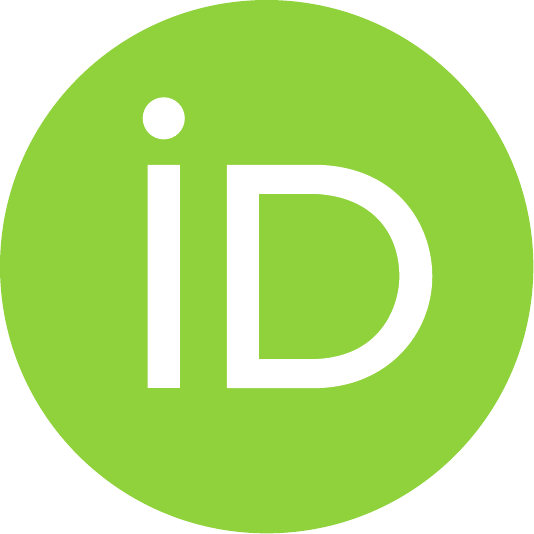}}\hspace*{1pt}$ and  Ricardo Villanueva-Polanco$^{2}\href{https://orcid.org/0000-0002-8682-4830} {\includegraphics[height=\fontcharht\font`B]{orcid.pdf}}\hspace*{1pt}$}
\institute{$^{1}$Inria and Laboratoire d’Informatique de l’Ecole polytechnique,\\
Institut Polytechnique de Paris, Palaiseau, France \\ \texttt{gustavo@cryptme.in}\\$^{2}$Technology Innovation Institute, UAE \texorpdfstring{\\}{, }  \texttt{ ricardo.polanco@tii.ae}}
\date{March 2024}
}
\begin{document}

\maketitle
\begin{abstract}
SNOVA is a post-quantum cryptographic signature scheme known for its efficiency and compact key sizes, making it a second-round candidate in the NIST post-quantum cryptography standardization process. This paper presents a comprehensive fault analysis of SNOVA, focusing on both permanent and transient faults during signature generation. We introduce several fault injection strategies that exploit SNOVA's structure to recover partial or complete secret keys with limited faulty signatures. Our analysis reveals that as few as $22$ to $68$ faulty signatures, depending on the security level, can suffice for key recovery. We propose a novel fault-assisted reconciliation attack, demonstrating its effectiveness in extracting the secret key space via solving a quadratic polynomial system. Simulations show transient faults in key signature generation steps can significantly compromise SNOVA’s security. To address these vulnerabilities, we propose a lightweight countermeasure to reduce the success of fault attacks without adding significant overhead. Our results highlight the importance of fault-resistant mechanisms in post-quantum cryptographic schemes like SNOVA to ensure robustness.
\end{abstract}

\begingroup
\makeatletter
\def\@thefnmark{} \@footnotetext{\relax
Author list in alphabetical order; see
\url{https://www.ams.org/profession/leaders/culture/CultureStatement04.pdf}.
\def\ymdtoday{\leavevmode\hbox{\the\year-\twodigits\month-\twodigits\day}}\def\twodigits#1{\ifnum#1<10 0\fi\the#1}%
}
\endgroup

\section{Introduction}
\label{sec:intro}
The National Institute of Standards and Technology (NIST) initiated an additional call for post-quantum digital signature proposals to introduce variability in the mathematical foundations of digital signatures. In response, NIST received 40 submissions based on diverse mathematical problems. Among these, 10 submissions were based on multivariate polynomial equations over finite fields, a branch of post-quantum cryptography known as MQ-based cryptography.

MQ-based cryptography relies on the difficulty of solving systems of multivariate quadratic equations over finite fields. The fundamental problem can be defined as follows: given a system of \( m \) quadratic equations in \( n \) variables over a finite field \( \mathbb{F}_q \), find a solution \( \mathbf{x} = (x_1, x_2, \ldots, x_n) \in \mathbb{F}_q^n \) such that:

\[
\begin{cases}
    Q_1(x_1, x_2, \ldots, x_n) = 0 \\
    Q_2(x_1, x_2, \ldots, x_n) = 0 \\
    \vdots \\
    Q_m(x_1, x_2, \ldots, x_n) = 0
\end{cases}
\]where each \( Q_i \) is a quadratic polynomial of the form:

\[
Q_i(x_1, x_2, \ldots, x_n) = \sum_{1 \leq j \leq k \leq n} a_{ijk} x_j x_k + \sum_{1 \leq j \leq n} b_{ij} x_j + c_i,
\]with coefficients \( a_{ijk}, b_{ij}, c_i \in \mathbb{F}_q \).

The security of MQ-based cryptography is based on the computational hardness of the Multivariate Quadratic problem. Specifically, for large values of \( n \), solving a random system of such equations is known to be NP-hard, making it computationally infeasible for an attacker to solve within a reasonable time frame, even with powerful computational resources.

In addition to the inherent mathematical complexity, implementing robust protections is essential for securing MQ-based cryptographic schemes against both side-channel and fault attacks. Side-channel attacks exploit various forms of leakage—such as timing variations, power consumption, or electromagnetic emissions—to gain insights into the cryptographic process. These attacks take advantage of unintended information leaks that arise during the physical implementation of a cryptographic algorithm, rather than exploiting weaknesses in the algorithm itself.

Fault attacks, in contrast, involve deliberately introducing errors during cryptographic execution, such as memory corruption or inducing bit flips, to extract sensitive information by analyzing erroneous outputs. To mitigate these threats, countermeasures such as constant-time algorithms, masking techniques, noise generation, and redundancy checks can be integrated, significantly enhancing the robustness of cryptographic implementations. These safeguards help ensure that the theoretical security of MQ-based schemes translates into practical resilience, providing strong protection against both side-channel and fault attacks in real-world environments.

\subsection{MQ signature schemes}
One of the earliest efforts to convert the MQ problem into a digital signature scheme was introduced in 1988 with the \(C^*\) scheme~\cite{c_star}. However, Patarin successfully attacked this scheme in 1995~\cite{Patarin95}, rendering it insecure. Since then, numerous advancements have been made in digital signature schemes based on multivariate polynomials. A prominent example of these advancements is the Unbalanced Oil-Vinegar (UOV) scheme, which offers strong security features.

The UOV-like schemes are an extension of the original Oil-Vinegar (OV) scheme and they are designed to enhance security by introducing an imbalance between the number of ``oil'' and ``vinegar'' variables. We can briefly define UOV as: let \( v \) be the number of vinegar variables: \( v_1, v_2, \ldots, v_v \), and \( o \) be the number of oil variables: \( o_1, o_2, \ldots, o_o \).
The public key consists of \( o \) quadratic polynomials \( P_i \) in \( n = v + o \) variables over a finite field \( \mathbb{F}_q \), defined as:
\begin{align*}
P_i(v_1, \ldots, v_v, o_1, \ldots, o_o) &= \sum_{1 \leq j \leq k \leq v} a_{ijk} v_j v_k + \sum_{j=1}^v \sum_{k=1}^o b_{ijk} v_j o_k \\
&+ \sum_{1 \leq j \leq k \leq o} c_{ijk} o_j o_k + \sum_{j=1}^v d_{ij} v_j + \sum_{k=1}^o e_{ik} o_k + f_i,
\end{align*}
where \( a_{ijk}, b_{ijk}, c_{ijk}, d_{ij}, e_{ik}, f_i \in \mathbb{F}_q \).

The private key consists of the secret lineal transformation and the so-called central map that link the vinegar and oil variables to the public key polynomials. During the signing process, random values are selected for the vinegar variables \( v_1, \ldots, v_v \). These values are then substituted into the quadratic polynomials, resulting in a system of linear equations involving the oil variables. Next, the system is solved to determine the values of the oil variables \( o_1, \ldots, o_o \), and the signature is produced by combining the values of both the vinegar and oil variables.

To verify the signature, the signed values are substituted into the public polynomials to ensure they satisfy the equations \( P_i -y_i=0\) for all \( i \).

In the world of UOV-like schemes, we can remark that there are several benefits such as \textit{small} signatures, fast verification, and reasonable public key sizes. This is the case of MAYO~\cite{mayo, MAYO21} and SNOVA~\cite{SNOVA, cryptoeprint:2022/1742}.

\subsection{Fault attacks}

Fault attacks are techniques used to exploit cryptographic implementations. This section provides an overview of fault attacks.

\paragraph{Fault Attacks.}
Fault injection attacks intentionally disrupt a device's normal operation to induce errors and extract information from cryptographic processes. Techniques such as electromagnetic pulses, lasers, voltage glitches, and DRAM row hammering are commonly used for fault injection, each varying in precision and complexity~\cite{biham1997differential, agoyan2010efficient, kim2014flipping}. For example, laser-based injections offer high accuracy but are expensive, whereas DRAM row hammering requires extensive profiling. Inducing faults often requires multiple attempts; however, this does not necessarily mean causing multiple distinct faults. It may involve repeatedly inducing a specific fault over several runs to collect sufficient data for exploitation.

\paragraph{Fault Attacks against other MQ-based cryptosystems}
\begin{table}[htp]
\small
\centering
\caption{Comparison of Previous Fault Attacks on Multivariate Signature Schemes.}
\label{tab:previous_work}
\resizebox{\textwidth}{!}{%
\begin{tabular}{@{}l|c|c|c|c@{}}
\toprule
\textbf{Algorithm}      & \textbf{\#Signatures} & \textbf{\#Faults} & \textbf{Evaluation} & \textbf{Assumptions} \\ \midrule
Multiple~\cite{hashimoto2011}    & Multiple              & Multiple          & Theoretical         & None                \\ \midrule
UOV/Rainbow~\cite{kramer2019} & Multiple              & Multiple          & Theoretical         & None                \\ \midrule
UOV~\cite{shim2009}        & 44--103               & Multiple          & Theoretical         & None                \\ \midrule
LUOV~\cite{mus2020}         & Multiple              & Multiple          & Practical           & Key in $\mathbb{F}_2$        \\ \midrule
Rainbow~\cite{aulbach2022}       & Multiple              & 1                 & Simulation          & Exact memory reuse  \\ \midrule
UOV~\cite{furue2022}       & Multiple              & 2--40             & Simulation          & Enumeration $2^{41}$--$2^{89}$ \\ \midrule
MAYO~\cite{sayari2021}     & 2                    & 1                 & Theoretical         & None                \\ \midrule
MAYO~\cite{aulbach2023}       & 1                    & 1                 & Practical           & Zero-initialization  \\ \midrule
MAYO~\cite{jendral2024p}               & 1                    & 1                 & Practical, Simulation & None                \\\midrule
\textbf{SNOVA}  &  &  & & \\
permanent fault strategy & 22--68 & 1 & Theoretical, Simulation & None \\\midrule
\textbf{SNOVA}  &  &  &  &  \\
Fault-assisted reconciliation attack & 1 & Multiple & Theoretical, Simulation & None\\\bottomrule
\end{tabular}
}
\end{table}

Table~\ref{tab:previous_work} compares previous fault injection attacks on multivariate signature schemes, highlighting key features such as the number of signatures and faults required, the evaluation method, and any assumptions made.

Recently, \cite{jendral2024p} presented an attack on a MAYO implementation that successfully recovered the private key. This attack targeted a single execution of MAYO. However, the fault was not in MAYO itself, but rather in the C implementation of Keccak, which is responsible for generating the Vinegar and Oil variables. The paper exploits a vulnerability in the pseudorandom function, using a fault in this component to reveal information leading to private key recovery.

In this work, we explore a similar attack. However, instead of targeting Keccak, we focus directly on the vinegar variables by inducing faults—i.e., fixing specific bit values—which allows us to recover the private key. Despite this similarity, our approach differs in the algorithm used for key recovery. Moreover, we introduce an SNOVA fault-assisted reconciliation attack that requires only a single signature. This approach is distinct from previous works.

\subsection{Our contributions} 

In this paper, we explore fault injection attacks and their impact on the security of SNOVA. We present an analysis demonstrating that by inducing either a single permanent fault or transient faults in specific operations of the signature generation algorithm, an attacker can create faulty signatures that reveal partial private key information. Our comprehensive analysis covers several scenarios. First, we investigate which rows of $T$, the matrix representation of the SNOVA private linear map, an attacker can recover after fixing certain $\mathbb{F}_{16}$ elements in some (or all) vinegar variables during signature generation. We then generalize this by examining the recovery of rows of $T$ after fixing bits in the binary representation of the vinegar variables. In both cases, we show that collecting enough faulty signatures allows an attacker to recover $T$ partially, compromising the private key. We then turn our attention to another scenario where, by inducing transient faults in the generation of vinegar variables, the attacker can recover the secret space via a reconciliation attack. Simulations of our fault attack are included to substantiate our claims.

Our detailed examination highlights the critical need for fault attack countermeasures to implement SNOVA. Therefore, we also provide a countermeasure and its corresponding analysis to counteract our fault attacks. Finally, our findings underscore the importance of incorporating comprehensive security strategies to protect against this type of attack, ensuring the integrity and reliability of cryptographic systems.

\paragraph{Paper organization.} This paper is organized as follows. \cref{sec:snova} presents a self-contained description of the SNOVA signature scheme. \cref{sec:fault} presents our fault analysis on SNOVA, introducing several scenarios along with the implications of our recovery algorithms. \cref{sec:simulations} describes our experimental setting, our simulation of our fault attacks on SNOVA, and its results. \cref{sec:countermeasure} presents a countermeasure to protect SNOVA against the fault attack introduced previously. Finally, \cref{sec:conclusion} concludes our work by highlighting our findings and their implications and delineating future works.

\section{A simple non-commutative UOV scheme}
\label{sec:snova}
In this section, we introduce SNOVA, a simple non-commutative UOV scheme. SNOVA is a recently proposed UOV-like signature scheme, as outlined in~\cite{cryptoeprint:2022/1742}, and has been submitted to the NIST competition for Post-Quantum Digital Signature Schemes.

\noindent Let $v, o, l \in 
\mathbb{N}$ with $v > o$ and $\mathbb{F}_q$ a finite field with $q$ being a power of a prime number. Set $n = v+o$ and $m = o$. By $[m]$ we denote the set $\{1,\ldots,m\}$ and by $\mathcal{R}$ we denote the ring of $l \times l$ matrices over the finite field $\mathbb{F}_q$. Also, by $U = (U_1,\ldots ,U_n)^t \in \mathcal{R}^n$ we denote a column vector with $n$ entries from $\mathcal{R}$. Let $Q \in \mathcal{R}$, we denote by $\Lambda_{Q}$,  the $nl \times nl$ block diagonal matrix with $Q$-blocks along the diagonal.


\paragraph{The subring $\mathbb{F}_{q}[S]$ of $ \mathcal{R}$} is defined to be $$\mathbb{F}_{q}[S]=\{a_0+a_1S+\ldots+a_{l-1}S^{l-1}|a_0,a_1,\ldots,a_{l-1} \in \mathbb{F}_{q}\}$$ where $S$ is a $l \times l$ symmetric matrix with
irreducible characteristic polynomial. Note that the elements in $\mathbb{F}_{q}[S]$ are symmetric and all commute. Additionally, the non-zero elements in $\mathbb{F}_{q}[S]$ are invertible. In particular,  $\mathbb{F}_{q}[S]$  is a finite field with  $\mathbb{F}_{q}[S] \cong \mathbb{F}_{q^l}$.

\paragraph{The central map} is given by 
$\mathcal{F}=[\mathcal{F}_1, \mathcal{F}_2, \ldots, \mathcal{F}_m]: \mathcal{R}^{n} \rightarrow \mathcal{R}^{m}$, and $\mathcal{F}_i$ is defined as 

$$\mathcal{F}_k(X_1, X_2,  \ldots, X_n)=\sum_{\alpha =1 }^{l^2} A_{\alpha} \cdot \big (\sum_{(i,j) \in \Omega} X_i^ {t} \cdot (Q_{\alpha1}F_{k,ij}Q_{\alpha2}) X_j \big ) \cdot B_{\alpha}$$ where $\Omega  = \{(i, j) : 1 \leq i, j \leq n\} \setminus \{(i, j) : m + 1 \leq i, j \leq n\}$, $F_{k,ij} \xleftarrow{\$} \mathcal{R}$, $A_{\alpha},B_{\alpha} \xleftarrow{\$} \mathcal{R}$ (invertibles), and $Q_{\alpha 1},Q_{\alpha 2} \xleftarrow{\$} \mathbb{F}_{q}[S] $ (invertibles). Set $F_k=[F_{k,ij}]_{(i,j) \in \Omega}$ for each $k \in [m]$. 
\paragraph{Invertible linear map for this scheme} is the map $\mathcal{T} : \mathcal{R}^n \rightarrow \mathcal{R}^n$ corresponding to the matrix

    $$
T= \begin{bmatrix}
I^{11} & T^{12} \\
 O & I^{22}
\end{bmatrix}
$$ where $T^{12}$ is a $v \times o$ matrix consisting of nonzero entries $T^{12}_{i,j}$  chosen randomly from $\mathbb{F}_{q}[S]$. Note that $T^{-1} = T$, if $\mathbb{F}_{q}$ is of characteristic $2$. In addition, $I^{11}$ and $ I^{22}$ are identity matrices over $\mathcal{R}$ of size $v \times v$ and $o \times o$ respectively.

\paragraph{Public map} is defined as $\mathcal{P}:= \mathcal{F}\circ \mathcal{T}=[\mathcal{P}_1=\mathcal{F}_1 \circ \mathcal{T}, \ldots, \mathcal{P}_m=\mathcal{F}_m \circ \mathcal{T} ]$. Set $U = (U_1,\ldots ,U_n)^t \in \mathcal{R}^n$, then 

\begin{equation}\label{eq:P_i}
\mathcal{P}_k(U)=\sum_{\alpha=1}^{l^2}  A_{\alpha} (TU)^t\Lambda_{Q_{\alpha1}}F_{k}\Lambda_{Q_{\alpha2}}(TU)\cdot B_{\alpha} 
\end{equation}
\noindent for any $k \in [m]$. Moreover, $\mathcal{P}_k (U)$ can be written as

$$
\mathcal{P}_k (U)=\mathcal{F}_k(\mathcal{T} (U))=  \sum_{\alpha=1}^{l^2} \sum_{i=1}^{n} \sum_{j=1}^{n} A_{\alpha}\cdot U^t_{i}(Q_{\alpha1}P_{k,ij}Q_{\alpha 2})U_{j}\cdot B_{\alpha},
$$ where $P_{k, ij}=\sum_{(s,t) \in \Omega} T_{is} F_{k, st} T_{tj}$, by the commutativity of $\mathbb{F}_{q}[S]$ and that the elements in $\mathbb{F}_{q}[S]$ are symmetric. Set $P_k=[P_{k,ij}]_{(i,j) \in [n]\times[n]}$ for each $k \in [m]$.

The SNOVA signature scheme \cite{cryptoeprint:2022/1742} consists of a triple of algorithms $(\texttt{KeyGen}$, $\texttt{Sign}$, $\texttt{Verify}).$ Moreover, a SNOVA parameter set is given by values for $v,o, l, \lambda$.

The \texttt{KeyGen} function runs a probabilistic algorithm as shown by Algorithm \ref{scheme:keygen}. It outputs a SNOVA key pair $(\texttt{sk}, \texttt{pk})$. 

The public key \texttt{pk} is a representation of $\mathcal{P}$. A full public key consists of the list of matrices $\big \{P_k = \begin{bmatrix}
P_k^{11} & P_k^{12} \\
P_k^{21} & P_k^{22}
\end{bmatrix}
: k \in [m] \big \}$ and the list of matrices $\big \{ A_{\alpha}, B_{\alpha}, Q_{\alpha1}, Q_{\alpha2} : \alpha \in [l^2] \big \}$. However, it is enough to store $(\texttt{Spublic}, \{P_k^{22}\}_{k\in [m]})$. Indeed, the public seed $\texttt{Spublic}$ is used to regenerate $P_k^{11}$,  $P_k^{12}$ and $P_k^{21}$ for $k \in [m]$, and $A_{\alpha}, B_{\alpha}, Q_{\alpha1}$ and $Q_{\alpha2}$ for $\alpha \in \{1,\ldots ,l^2\}$. Therefore, the public key size is
$ m \cdot m^2 \cdot l^2$ field elements plus the size of
the public seed $\texttt{Spublic}$.

The private key \texttt{sk} is a representation of $(\mathcal{F}, \mathcal{T} )$. A full private key  consists of a matrix $T^{12}$ and the list of matrices $\big \{ F_k: k \in [m] \big \}$.  In practice, a private seed $\texttt{Sprivate}$ is used to generate $T^{12}$, and the matrices $\{F_k\}_{k\in [m]}$ are computed by exploiting the relation between $F_k$, $P_k$ along with $T^{12}$, as shown by the lines from \ref{kg1} to \ref{kg4} of Algorithm~\ref{scheme:keygen}.

Table~\ref{tab:parameters} summarizes current SNOVA parameters, and public and private keys sizes, as well as, signature sizes for each security level. 

\begin{algorithm}[htp]
\caption{Key generation algorithm}
\label{scheme:keygen}
\begin{algorithmic}[1]

\STATE \textbf{function} $\texttt{KeyGen}(v,o, l, \lambda)$
 \STATE $m\gets o$;$n\gets o+v$;
 \STATE $\texttt{Sprivate} \xleftarrow{\$} \{0,1\}^{2\lambda}$
 \STATE $\texttt{Spublic} \xleftarrow{\$} \{0,1\}^{2\lambda}$
 \STATE $T^{12} \gets \texttt{PRG} (\texttt{Sprivate})$, where $T_{ij}^{12} \in \mathbb{F}_{16}[S]$.

 \STATE $\{P_k^{11}\}_{k\in [m]},\{P_k^{12}\}_{k\in [m]},\{P_k^{21}\}_{k\in [m]}, \{A_{\alpha}\}_{\alpha\in [l^2]}, \{B_{\alpha}\}_{\alpha\in [l^2]}, \{Q_{\alpha1}\}_{\alpha\in [l^2]}, \{Q_{\alpha2}\}_{\alpha\in [l^2]} \gets \texttt{PRG} (\texttt{Spublic})$ 

 \FOR{($k\gets 1,k\leq m,k\gets k+1$)}

    \STATE \label{kg1}$F_k^{11} \gets P_k^{11}$
    \STATE \label{kg2} $F_k^{12}  \gets P_k^{11} T^{12} + P_k^{12}  $
    \STATE \label{kg3}$F_k^{21}  \gets (T^{12})^{t} P_k^{11}  + P_k^{21}  $
     \STATE\label{kg4} $P_k^{22}  \gets (T^{12})^{t} \big ( P_k^{11} T^{12}  + P_k^{12} \big ) + P_k^{21}T^{12} $

  \ENDFOR

  \STATE $\texttt{sk}\gets (\{F_k^{11}\}_{k\in [m]}, \{F_k^{12}\}_{k\in [m]}, \{F_k^{21}\}_{k\in [m]}, T^{12})$
  \STATE $\texttt{pk}\gets (\texttt{Spublic}, \{P_k^{22}\}_{k\in [m]})$
  
\RETURN $(\texttt{sk},\texttt{pk})$;
\STATE \textbf{end function} 
\end{algorithmic}
\end{algorithm}

\begin{table}[h]
\centering
\caption{Parameters for SNOVA~\cite{SNOVA}.}
\begin{tabular}{c|c|c|c|c}
\toprule
Sec. Level & $(v, o, q, l, \lambda)$ & Public key (\si{\byte}) & Signature (\si{\byte})  & Private key (\si{\byte}) \\ 
\midrule
\multirow{3}{*}{I}   & $(37, 17, 16, 2, 128)$ & $9826(+16)$  & $108(+16)$    & $60008(+48)$    \\
                     & $(25, 8, 16, 3,128)$  & $2304(+16)$  & $148.5(+16)$  & $37962(+48)$    \\
                     & $(24, 5, 16, 4,128)$  & $1000(+16)$  & $232(+16)$    & $34112(+48)$    \\ 
\midrule
\multirow{3}{*}{III} & $(56, 25, 16, 2,192)$ & $31250(+16)$ & $162(+16)$    & $202132(+48)$   \\
                     & $(49, 11, 16, 3, 192)$ & $5989.5(+16)$& $270(+16)$    & $174798(+48)$    \\
                     & $(37, 8, 16, 4, 192)$  & $4096(+16)$  & $360(+16)$    & $128384(+48)$    \\ 
\midrule
\multirow{3}{*}{V}   & $(75, 33, 16, 2, 256)$ & $71874(+16)$ & $216(+16)$    & $515360(+48)$    \\
                     & $(66, 15, 16, 3, 256)$ & $15187.5(+16)$& $364.5(+16)$ & $432297(+48)$    \\
                     & $(60, 10, 16, 4, 256)$ & $8000(+16)$  & $560(+16)$    & $389312(+48)$    \\ 
\bottomrule
\end{tabular}

\label{tab:parameters}
\end{table}

The \texttt{Sign} function runs a UOV-like signing procedure as shown by Algorithm \ref{scheme:sign}. It digitally signs a message $\texttt{M}$ under the private key $\texttt{sk}$. It first samples a $\texttt{salt}$ from $\{0,1\}^{2\lambda}$, then sets 
$Y \gets \mathcal{H}_1(\texttt{Spublic} ||\mathcal{H}_0(\texttt{M})||  \texttt{salt})$ where $ Y \in \mathcal{R}^m$. The algorithm then chooses random values $V_1,\ldots ,V_v \in \mathcal{R}$ as the vinegar variables. Then, it attempts to find the values $(V_{v+1}, \ldots , V_{n})$ by solving  the equation $\mathcal{F}(V_1,\ldots ,V_v,V_{v+1},\ldots ,V_n) = Y$. If no solution to the equation is found, the algorithm will choose random values $V'_1,\ldots ,V'_v \in \mathcal{R}$ and repeat the procedure until it finds a solution to the equation. Let $X = (V_1, \ldots , V_v, V_{v+1}, \ldots , V_n)^{t}$ be the solution to the equation.  This algorithm then computes the signature as $\texttt{S} = \mathcal{T}^{-1}(X)$ and outputs $(\texttt{S} , \texttt{salt})$.

\begin{algorithm}[htp] 
\scriptsize
\caption{signs message $\texttt{M}$}
\label{scheme:sign}
\begin{algorithmic}[1]

\STATE \textbf{function} $\texttt{sign}(v,o, l, \lambda, \texttt{sk}, spublic, \texttt{M})$
\STATE $m\gets o$
\STATE $n\gets o+v$
\STATE \label{pkexpansion}$(\{F_k^{11}\}_{k\in [m]}, \{F_k^{12}\}_{k\in [m]}, \{F_k^{21}\}_{k\in [m]}, T^{12}) \gets \texttt{sk} $
\STATE $ \{A_{\alpha}\}_{\alpha\in [l^2]}, \{B_{\alpha}\}_{\alpha\in [l^2]}, \{Q_{\alpha1}\}_{\alpha\in [l^2]}, \{Q_{\alpha2}\}_{\alpha\in [l^2]} \gets PRG (\texttt{spublic})$ 
\STATE $\texttt{digest} \gets \mathcal{H}_0(\texttt{M})$
\STATE $\texttt{salt}\xleftarrow{R}\{0,1\}^{\lambda}$

\STATE $\texttt{is\_done} \gets \texttt{False}$
\STATE $ cont \gets 0$;

\STATE $[Y_1,Y_2,\ldots, Y_{m} ] \gets \mathcal{H}_1(\texttt{Spublic} ||\texttt{digest}  || \texttt{salt})$

\STATE $F_{k}\gets\sum_{\alpha=1}^{l^2}A_{\alpha}\big (\sum_{i=1}^{v}\sum_{j=1}^{v}X_{i}^t(Q_{\alpha1}F^{11}_{k,ij}Q_{\alpha2})X_j +\sum_{i=1}^{v}\sum_{j=1}^{m}X_i^t(Q_{\alpha1}F^{12}_{k,ij}Q_{\alpha2})X_j+\sum_{j=1}^{m}\sum_{i=1}^{v}X_j^t(Q_{\alpha1}F^{21}_{k,ji}Q_{\alpha2})X_i \big )\cdot B_{\alpha}
$ 

for all $i \in [m]$.
\WHILE{\texttt{not} $\texttt{is\_done}$}
\STATE \label{compvin1} $[\texttt{V}_1,\texttt{V}_2, \ldots, \texttt{V}_v] \gets PRG(Sprivate|| \texttt{digest}||\texttt{salt} || cont)$

\STATE \label{compvin2} Compute $F_{k,VV} \gets \sum_{\alpha=1}^{l^2}A_{\alpha}\big (\sum_{i=1}^{v}\sum_{j=1}^{v}\texttt{V}_i^t(Q_{\alpha1}F^{11}_{k,ij}Q_{\alpha2})\texttt{V}_j \big )\cdot B_{\alpha}
$ for all $k \in [m]$.

\STATE Express $Y_k - F_{k,VV} =\sum_{\alpha=1}^{l^2}A_{\alpha}\big (\sum_{i=1}^{v}\sum_{j=1}^{m}\texttt{V}_i^t(Q_{\alpha1}F^{12}_{k,ij}Q_{\alpha2})X_j+\sum_{j=1}^{m}\sum_{i=1}^{v}X_j^t(Q_{\alpha1}F^{21}_{k,ji}Q_{\alpha2})\texttt{V}_i \big )\cdot B_{\alpha}
$ for all $k \in [m]$ as an equation system on the oil variables $\overrightarrow{X_1}, \overrightarrow{X_2},  \ldots, \overrightarrow{X_{m}}$.

 \medskip

$\begin{matrix}
M_{1,1}\overrightarrow{X_1}+M_{1,2}\overrightarrow{X_{1}}+\ldots +M_{1,m}\overrightarrow{X_{m}} =\overrightarrow{Y_1}-\overrightarrow{F_{1,VV}} \\
M_{2,1}\overrightarrow{X_1}+M_{2,2}\overrightarrow{X_{2}}+\ldots +M_{2,m}\overrightarrow{X_{m}} =\overrightarrow{Y_2}-\overrightarrow{F_{2,VV}} \\ 
\vdots \\ 
M_{m,1}\overrightarrow{X_1}+M_{m,2}\overrightarrow{X_{2}}+\ldots+ M_{m,m}\overrightarrow{X_{m}} =\overrightarrow{Y_{m}}-\overrightarrow{F_{m,VV}}

\end{matrix}$

\STATE Represent this equation system as an $(ml^2) \times (ml^2+1)$ matrix $\mathbf{A}$ over $\mathbb{F}_{16}$
\STATE $ L_{O}, \texttt{output}  \gets \texttt{Gauss}(\mathbf{A})$

\IF{$\texttt{output}$}
  
  \STATE $[\texttt{V}_{v+1},\texttt{V}_{v+2}, \ldots, \texttt{V}_{n}] \gets L_{O}$
  \STATE $\texttt{V} \gets [\texttt{V}_1,\texttt{V}_2, \ldots, \texttt{V}_v, \texttt{V}_{v+1}, \ldots, \texttt{V}_{n}]$
  \STATE \label{sigcom} $T \gets \begin{bmatrix}
I^{11} & T^{12} \\
 O & I^{22}
\end{bmatrix} $
  \STATE \label{sig:applyT}$\texttt{S} \gets T \cdot \texttt{V}^t$
   \STATE $\texttt{is\_done} \gets \texttt{True}$
   
\ELSE

\STATE $cont \gets cont+1$
\ENDIF

\ENDWHILE
 
\RETURN $(\texttt{S},\texttt{salt})$;
\STATE \textbf{end function} 
\end{algorithmic}
\end{algorithm}

The \texttt{Verify} function runs a deterministic algorithm. It simply verifies if a signature $(\texttt{S},\texttt{salt})$ for $\texttt{M}$ is valid under the public key $\texttt{pk}$. If $\mathcal{H}_1(\texttt{Spublic}$$||\mathcal{H}_0(\texttt{M})$ $||\texttt{salt}) = \mathcal{P}(\texttt{S})$, then the signature is accepted, otherwise it is rejected.

\subsubsection{Reconciliation attack.}
\label{recon_attack_sec}
Ikematsu, and Akiyama~\cite{cryptoeprint:2024/096}, Li and Ding~\cite{cryptoeprint:2024/110}, and Nakamura, Tani, and Furue~\cite{cryptoeprint:2024/1374} analyzed the security of SNOVA against key-recovery attacks, unveiling all known key-recovery attacks for an instance of SNOVA can be seen as key-recovery attacks to instances of an equivalent UOV signature scheme. Particularly, \cite{cryptoeprint:2024/096} and  \cite{cryptoeprint:2024/110} concluded that all known key-recovery attacks for SNOVA with parameters $(v,o,l,q)$ can be seen as attacks to a UOV signature scheme with $lo^2$ equations and $l(v+o)$ variables over $\FF_{q}$. 
In particular, for the \textit{reconciliation} attack, the attacker must find a specific solution $\mathbf{u}_0 \in \mathbb{F}_q^{ln}$ from among many solutions of a quadratic polynomial system of the form

\begin{equation}
   \label{rec} \mathbf{u}_0^t(\Lambda_{S^{i}}P_k\Lambda_{S^{j}})\mathbf{u}_0=0 \in \mathbb{F}_q,
\end{equation}for~$k \in [m],i,j \in \{0,1,\ldots, l-1\}$. 

Once $\mathbf{u}_0$ is found, any $\mathbf{u}$ in the linearly independent set $\{\Lambda_{S^{j}}\mathbf{u}_0: 0 \leq j\leq l-1 \}$ will also satisfy \cref{rec}. Additionally, the remaining vectors in the secret space $\mathcal{O}$ can be determined by leveraging the fact that for any $\mathbf{u},\mathbf{v} \in \mathcal{O}$, it holds
\begin{equation}
   \label{a} \mathbf{v}^t(\Lambda_{S^{i}}P_k\Lambda_{S^{j}})\mathbf{u}=0 \in \mathbb{F}_q ~\textrm{for}~k \in [m], 0\leq i,j \leq l-1.
\end{equation} 

Finally, for any $U \in \mathcal{K}$, it holds $\mathcal{P}_k(U )=0$ for all $k\in [m]$, where 
   \begin{equation}
   \label{secret_space_R} 
   \mathcal{K}:=\mathcal{O} \otimes \mathbb{F}_q^{l} = \{ \mathbf{u} \otimes \mathbf{e}^t \in \mathcal{R}^n: \mathbf{u} \in \mathcal{O}, \mathbf{e}\in \mathbb{F}_q^{l}   \}. 
\end{equation}

Thus, the complexity of the \textit{reconciliation} attack is dominated by finding a solution to the quadratic system in \cref{rec}. A recent paper \cite{cryptoeprint:2024/1770} introduces a new algorithm that exploits the stability of the quadratic system in \cref{rec} under the action of a commutative group of matrices, reducing the complexity of solving SNOVA systems, over generic ones. In particular, they show how their new algorithm decreases the complexity of solving such a system. On the other hand, we here explore other directions by introducing a new fault-assisted reconciliation attack in~\cref{sec:Fault_many}. This attack leverages induced transient faults to recover the secret key space by solving the system as mentioned earlier.


\section{Fault analysis on SNOVA}
\label{sec:fault}

In this section, we present our comprehensive fault analysis on SNOVA. We first state our attack model, then explore a first unsuccessful attempt at mounting a fault attack, and then turn our attention to our successful attempts at mounting a fault attack against SNOVA.

\subsection{Adversarial Model}
\label{sec:adversary_model}
We consider an adversarial model similar to those assumed in previous works, particularly in~\cite{kramer2019}, within the context of UOV and RAINBOW.

In this model, the attacker targets the signature generation process and can induce transient or permanent faults affecting specific operations within \cref{scheme:sign}. These faults enable the attacker to manipulate certain values during the execution of the signature generation algorithm. Importantly, the attacker may be unaware of the exact number of manipulated values or their specific content.

Subsequently, the attacker can invoke the faulty \cref{scheme:sign} multiple times --where the number of invocations depends on the attack strategy-- to gather message-signature pairs. Each signature is generated with the tampered values, and the attacker's ultimate goal is to analyze these pairs to extract partial information about the private key.

\subsection{Attack strategy by fixing field elements of the central map}

\begin{enumerate}
    \item The attacker causes a single permanent fault, which affects line \ref{pkexpansion} of \cref{scheme:sign}, such that some  $\mathbb{F}_{q}$ elements in $F_i \in \mathbb{F}_q^{ln\times ln}$, for $i \in I \subseteq [m]$ and $|I|\geq 1$, are fixed and unknown. In particular, for $F_i$, there is a fixed non-empty subset $J_i \subseteq [nl] \times [nl] $, such that  $\bar{F}_{i,(r_0,r_1)} \in \mathbb{F}_q$, $(r_0,r_1) \in J_i$ is fixed and unknown. We remark the line \ref{pkexpansion} in practice is an expansion of a private seed along with other operations to compute the list of matrices $\{F_k\}_{k\in[m]}$.

    \item For each $\omega \in [N_{msg}]$, the attacker calls \cref{scheme:sign} for the randomly chosen message 
$\texttt{M}^{(\omega)}  \in \mathcal{R}^{m}$ and receives the signatures $(\texttt{S}^{(\omega)}, \texttt{salt}^{(\omega)})$.
\end{enumerate}

Let $\bar{\mathcal{F}}$ be the faulty central map and $\bar{\mathcal{P}}=\bar{\mathcal{F}}\circ \mathcal{T}$ be the faulty public map. 
Recall that the SNOVA public map is defined as 

$$\mathcal{P}_k(\texttt{S}^{(\omega)})=\sum_{\alpha=1}^{l^2}  A_{\alpha} (T\texttt{S}^{(\omega)})^t\Lambda_{Q_{\alpha1}}F_{k}\Lambda_{Q_{\alpha2}}(T\texttt{S}^{(\omega)})\cdot B_{\alpha}$$ for any $k \in [m]$. Therefore,  it holds  

\begin{equation}
  \label{noisyeq}
     \bar{\mathcal{P}}_k(\texttt{S}^{(\omega)})-\mathcal{P}_k(\texttt{S}^{(\omega)}) = \sum_{\alpha=1}^{l^2}  A_{\alpha} (\texttt{V}^{(\omega)})^t\Lambda_{Q_{\alpha1}}(\bar{F}_{k}-F_{k})\Lambda_{Q_{\alpha2}}(\texttt{V}^{(\omega)})\cdot B_{\alpha}
\end{equation}

\noindent where $T\texttt{S}^{(\omega)}=\texttt{V}^{(\omega)}$ with $\texttt{V}^{(\omega)}=(\texttt{V}_1^{(\omega)},\ldots, \texttt{V}_n^{(\omega)})^t$, by the line \ref{sig:applyT} of \cref{scheme:sign}.

We remark the attacker can compute the left hand of \cref{noisyeq}, since $\bar{\mathcal{P}}_k(\texttt{S}^{(\omega)})=\mathcal{H}_1(\texttt{Spublic} ||\mathcal{H}_0(\texttt{M}^{(\omega)}) || \texttt{salt}^{(\omega)})_k$ and $\mathcal{P}$ is public.

Moreover, for any $i \in [m]\setminus I$, both sides of \cref{noisyeq} vanish. However, for any $i \in I$, the left hand of \cref{noisyeq} is expected to be a non-zero element in $\mathcal{R}$, and  $\tilde{F}_{i}=\bar{F}_{i}-F_{i} \in  \mathbb{F}_q^{ln\times ln}$, on the right side of \cref{noisyeq}, is expected to become a sparse matrix, since the entries $\tilde{F}_{i, st} \in \mathbb{F}_q$  with $(s,t) \in J_i$ are the only ones expected to be non-zero.  

We remark, nonetheless, that these observations may not be easily exploitable for the attacker to gain information on $T$, since the attacker does not know $I, \tilde{F}_{i}, J_i, \texttt{V}^{(\omega)}$ and $\bar{P}_i$. Therefore, our following scenario focuses on inducing a permanent fault affecting the line \ref{compvin1} of \cref{scheme:sign} to further exploit the relation $T\texttt{S}^{(\omega)}=\texttt{V}^{(\omega)}$.

\subsection{Attack strategy by fixing field elements  of vinegar variables}
\label{partial_recovery_field}
\begin{enumerate}
  \item The attacker introduces a single permanent fault, which affects line \ref{compvin1} of \cref{scheme:sign}, causing certain $\mathbb{F}_{q}$ elements in $\texttt{V}_i \in \mathcal{R}$, for $i \in I \subseteq [v]$ with $|I| \geq 1$, to be fixed and unknown. Specifically, for each variable $\texttt{V}_i$, there is a fixed non-empty subset $J_i \subseteq [l] \times [l]$, such that $\bar{\texttt{V}}_{i,(r_0, r_1)} \in \mathbb{F}_q$ for $(r_0, r_1) \in J_i$ is fixed and unknown.

    \item For each $\omega \in [N_{msg}]$, the attacker calls \cref{scheme:sign} for the randomly chosen message 
$\texttt{M}^{(\omega)}  \in \mathcal{R}^{m}$ and receives the signature 
$(\texttt{S}^{(\omega)}, \texttt{salt}^{(\omega)})$.
\item \label{recovery_field}The attacker then calls \cref{scheme:recover_from_partial} with parameters $v, o, l, [\texttt{S}^{(1)}, \ldots, \texttt{S}^{(N_{msg})}]$ to obtain $\texttt{dic}$, a dictionary-like data structure indexed by $[v] \times [l]^2$. When $N_{msg} > lo + 1$, \cref{scheme:recover_from_partial} will output \texttt{dic} such that $\texttt{dic}[(i, r_0, r_1)] = [T^{12}_{i1}, \ldots, T^{12}_{io}]$ for $i \in I$ and $(r_0, r_1) \in J_i$, and $\texttt{dic}[(i, r_0, r_1)] = \texttt{None}$ otherwise.
\end{enumerate}

\paragraph{Why is Step \ref{recovery_field} of the attack strategy expected to work correctly?}

As seen in \cref{sec:snova}, the invertible linear map $\mathcal{T}$ for the SNOVA scheme is given by the matrix $$T= \begin{bmatrix}
I^{11} & T^{12} \\
 O & I^{22}
\end{bmatrix},$$ where $T^{12}$ is a $v \times o$ matrix with nonzero entries $T^{12}_{ij}$  chosen randomly from $\mathbb{F}_{q}[S]$. From the line \ref{sig:applyT} of  \cref{scheme:sign}, it holds $T\texttt{S}^{(\omega)}=\texttt{V}^{(\omega)}$, that is,

$$ \begin{bmatrix}
1 & 0 & \ldots & 0&T^{12}_{11}&\ldots &T^{12}_{1o} \\
\vdots & \vdots & \ddots & \vdots&\vdots &\ddots &\vdots \\
0 & 0 & \ldots & 1 & T^{12}_{v1}&\ldots &T^{12}_{vo} \\
\vdots & \vdots & \ddots & \vdots&\vdots &\ddots &\vdots \\
0 & 0 & \ldots & \ldots &0 &\ldots&1 \\
\end{bmatrix}\begin{bmatrix}
\texttt{S}_1^{(\omega)}\\
\vdots\\
\texttt{S}_v^{(\omega)}\\
\vdots\\
\texttt{S}_n^{(\omega)}\\
\end{bmatrix}=\begin{bmatrix}
\texttt{V}_1^{(\omega)}\\
\vdots\\
\texttt{V}_v^{(\omega)}\\
\vdots\\
\texttt{V}_n^{(\omega)}\\
\end{bmatrix}.$$

Since $T^{12}_{ij} \in \mathbb{F}_{q}[S]$,  it holds $T^{1,2}_{ij}=\sum_{j_1=1}^{l}t^{12}_{ij,j_1}S^{j_1-1}$, where $t^{12}_{ij,j_1} \in \mathbb{F}_{q}$, and

\begin{align*} 
\texttt{S}^{(\omega)}_i  + \sum_{j=1}^{o}T^{12}_{ij}\texttt{S}^{(\omega)}_{v+j} =\texttt{S}^{(\omega)}_i+ \sum_{j=1}^{o}\sum_{j_1=1}^{l}t^{12}_{ij,j_1}S^{j_1-1}\texttt{S}^{(\omega)}_{v+j}&=\texttt{V}^{(\omega)}_{i}, i\in [v]  \\
\texttt{S}^{(\omega)}_{i} &=\texttt{V}^{(\omega)}_{i}, v+1\leq i\leq n
\end{align*} 

For $2 \leq \omega \leq N_{msg}$, we can write

\begin{align*} 
\texttt{S}^{(\omega)}_i-\texttt{S}^{(1)}_i +\sum_{j=1}^{o}\sum_{j_1=1}^{l}t^{12}_{ij,j_1}S^{j_1-1}(\texttt{S}^{(\omega)}_{v+j}-\texttt{S}^{(1)}_{v+j})&=\texttt{V}^{(\omega)}_{i} -\texttt{V}^{(1)}_{i}, i\in [v]  \\
\texttt{S}^{(\omega)}_{i}-\texttt{S}^{(1)}_{i} &=\texttt{V}^{(\omega)}_{i}-\texttt{V}^{(1)}_{i}, v+1\leq i\leq n
\end{align*}

\noindent Let us fix $2 \leq \omega \leq N_{msg}$ and $i\in [v]$. Also, let us set $\texttt{S}^{(\omega,1)}_i=\texttt{S}^{(\omega)}_i-\texttt{S}^{(1)}_i$,  $\texttt{S}^{(j_1,\omega,1)}_{v+j}=S^{j_1-1}(\texttt{S}^{(\omega)}_{v+j}-\texttt{S}^{(1)}_{v+j})$ and $\texttt{V}^{(\omega,1)}_{i}=\texttt{V}^{(\omega)}_{i}-\texttt{V}^{(1)}_{i}$. Consider 
\begin{equation}
\label{eq1}
\texttt{S}^{(\omega,1)}_i+ \sum_{j=1}^{o}\sum_{j_1=1}^{l}t^{12}_{ij,j_1}\texttt{S}^{(j_1,\omega,1)}_{v+j} =\texttt{V}^{(\omega,1)}_{i}. 
\end{equation}
Since \cref{eq1} is defined  over $\mathcal{R}$, it is equivalent to $l^2$ equations on $lo$ unknowns over $\mathbb{F}_q$. Therefore,  

\begin{equation}
\label{eq2}
    \texttt{S}^{(\omega,1)}_{i, (r_0, r_1)} +\sum_{j=1}^{o}\sum_{j_1=1}^{l}t^{12}_{ij,j_1} \texttt{S}^{(j_1,\omega,1)}_{v+j,(r_0, r_1)}=\texttt{V}^{(\omega,1)}_{i,(r_0, r_1)}, ~\textrm{for}~ r_0, r_1 \in [l].
\end{equation}

Note that the attacker can compute $\texttt{S}^{(\omega,1)}$ and $\texttt{S}^{(j_1,\omega,1)}$ on the left hand of \cref{eq1}. Additionally, for a fixed $i \in I$ and for $2\leq \omega \leq N_{msg}$,  a linear system of $(N_{msg} -1)|J_i|$ equations and $lo$ unknowns over $\mathbb{F}_q$ can be obtained and is given by \begin{equation}
\label{eq3}
    \{\texttt{S}^{(\omega,1)}_{i, (r_0, r_1)} +\sum_{j=1}^{o}\sum_{j_1=1}^{l}t^{12}_{ij,j_1} \texttt{S}^{(j_1,\omega,1)}_{v+j,(r_0, r_1)}=0, ~\textrm{for}~ (r_0, r_1) \in J_i \}_{2\leq \omega \leq N_{msg}}.
\end{equation}However the attacker does not know either $I$ or $J_i$. If $J_i$ were known by the attacker, collecting $N_{msg} > \frac{o\cdot l}{|J_i|}+1$ would be enough to guarantee an unique solution to the linear system of \cref{eq3}.

In order for the attacker to gain knowledge of $I$ and $J_i$ for $i 
\in I$, and recover $T^{12}
_{i,j}$ for $i \in I, j \in [o],$ the attacker may try to solve $v \cdot l^2$ linear systems separately, i.e. one for $i \in [v]$ and  $(r_0,r_1) \in [l]^2$, 

\begin{equation}
\label{eq4}
    \{\texttt{S}^{(\omega,1)}_{i, (r_0, r_1)} +\sum_{j=1}^{o}\sum_{j_1=1}^{l}t^{12}_{ij,j_1} \texttt{S}^{(j_1,\omega,1)}_{v+j,(r_0, r_1)}=0 \}_{2\leq \omega \leq N_{msg}}, 
\end{equation} where each has $N_{msg}-1$ equations and $lo$ unknowns. If $N_{msg}> l \cdot o+1$, then the linear systems for $(r_0,r_1)\in J_i, i \in I$ will have a unique solution, while the other linear systems are expected to have no solution. Therefore, when $N_{msg}> lo+1$, \cref{scheme:recover_from_partial} will output \texttt{dic} such that $\texttt{dic}[(i,r_0, r_1)]=[T^{12}_{i1},\ldots, T^{12}_{io}]$ for $i\in I, (r_0, r_1) \in J_i$ and $\texttt{dic}[(i,r_0, r_1)]=\texttt{None}$ otherwise.

Furthermore, if the attacker is able to fix at least an entry of each vinegar variable (i.e. $I=[v]$ and so $|J_i|\geq 1$) and collect at least $lo+2$ signatures,   \cref{scheme:recover_from_partial} will recover the entire matrix $T^{12}$.

\begin{algorithm}[htp]  
\caption{partially recovers $T^{12}$ from fixed field elements.}
\label{scheme:recover_from_partial}
\begin{algorithmic}[1]

\STATE \textbf{function} $\texttt{recover\_partial\_T\_F16}(v,o, l,[\texttt{S}^{(1)},\ldots, \texttt{S}^{(N_{msg})}])$
\STATE $S \gets \texttt{getS}(l)$
\STATE $\texttt{dic}\gets \{\}$

\FOR{ ($i\gets 1, i\leq v, i \gets  i+1$)}

\FOR{($r_0\gets 1, r_0\leq l, r_0 \gets  r_0+1)$}
\FOR{($r_1\gets 1, r_1\leq l, r_1 \gets  r_1+1)$}
\STATE $\mathbf{A} \gets \mathbb{F}_{16}^{(N_{msg}-1) \times ol}$
\STATE $\mathbf{Y} \gets \mathbb{F}_{16}^{(N_{msg}-1) \times 1}$
\FOR{ ($\omega\gets 2, \omega\leq N_{msg}, \omega \gets  \omega+1$)}
      \STATE $\texttt{S}^{(\omega,1)}_i \gets \texttt{S}^{(1)}_i-\texttt{S}^{(\omega)}_i$
         
                    \FOR{($j\gets 1, j\leq o, j \gets  j+1)$}
                       \FOR{($j_1 \gets 1, j_1\leq l, j_1 \gets  j_1+1)$}
                       \STATE $\texttt{S}^{(j_1,k,1)}_{v+j}  \gets S^{j_1-1}(\texttt{S}^{(\omega)}_{v+j}-\texttt{S}^{(1)}_{v+j})$
                       \STATE $\mathbf{A}_{(\omega-1),j\cdot l+j_1}  \gets \texttt{S}^{(j_1,k,1)}_{v+j,(r_0,r_1)}$
                      \ENDFOR
                    \ENDFOR
                  \STATE $\mathbf{Y}_{(\omega-1),0}\gets \texttt{S}^{(\omega,1)}_{i,(r_0,r_1)}$
                \ENDFOR             
            \STATE $ \texttt{X}, \texttt{output}  \gets \texttt{Gauss}(\mathbf{A}, \mathbf{Y} )$

\IF{$\texttt{output}$}
  
  \STATE $[T^{12}_{i1},\ldots, T^{12}_{io}] \gets \texttt{get\_elements\_in\_FqS}(\texttt{X},l)$
  \STATE $\texttt{dic}[(i, r_0,r_1)] \gets [T^{12}_{i1},\ldots, T^{12}_{io}] $
  
\ELSE

\STATE $\texttt{dic}[(i,r_0,r_1)] \gets \texttt{None} $
\ENDIF

            \ENDFOR  
 \ENDFOR

 \ENDFOR

 \RETURN $\texttt{dic}$
 
\STATE \textbf{end function} 
\end{algorithmic}
\end{algorithm}

\subsection{What if the attacker only can fix some bits of $V_i$ for $i \in I$?}\label{partial_recovery_bits}
In this section, we assume a variable $\texttt{V}_i$ is represented as a bit-string of length $N_{bits}\cdot l^2$, where $q=2^{N_{bits}}$ as it is the case for SNOVA. The attack strategy is as follows.

\begin{enumerate}
    \item The attacker causes a single permanent fault, which affects the line \ref{compvin1} of \cref{scheme:sign}, such that some bits of $\texttt{V}_i \in \mathcal{R}$, with $i \in I \subseteq [v]$ and $|I|\geq 1$, are fixed and unknown. In particular, for the variable $\texttt{V}_i$, there is a fixed non-empty subset $B_i \subseteq [l]\times[l]\times[N_{bits}]$, such that  $\bar{\texttt{V}}_{i,(r_0,r_1,b)} \in \mathbb{F}_2$, $(r_0,r_1,b) \in B_i$ is fixed and unknown.

    \item For each $\omega \in [N_{msg}]$, the attacker calls \cref{scheme:sign} for the randomly chosen message 
$\texttt{M}^{(\omega)}  \in \mathcal{R}^{m}$ and receives the signature $(\texttt{S}^{(\omega)}, \texttt{salt}^{(\omega)})$.

\item \label{recovery_bits} The attacker calls \cref{scheme:recover_from_partial_bits} with parameters $v,o,l,[\texttt{S}^{(1)},\ldots, \texttt{S}^{(N_{msg})}]$ to get $\texttt{dic}$, a dictionary-like data structure indexed by $[v]\times[l]^2\times[N_{bits}]$. When $N_{msg}> N_{bits} \cdot lo+1$, \cref{scheme:recover_from_partial_bits} will output \texttt{dic} such that $\texttt{dic}[(i,r_0, r_1,b)]=[T^{12}_{i1},\ldots, T^{12}_{io}]$ for $i\in I, (r_0, r_1,b) \in B_i$ and $\texttt{dic}[(i,r_0, r_1,b)]=\texttt{None}$ otherwise.
\end{enumerate}

\paragraph{Why is Step \ref{recovery_bits} of the attack strategy expected to work correctly?}

Since $\mathbb{F}_q$ can be seen as a vector space of dimension $N_{bits}$ over $\mathbb{F}_2$, we can obtain similar equations to those of \cref{eq3}. That is, for $i \in I$, we have

\begin{equation}
\label{eq5}
    \{\texttt{S}^{(\omega,1)}_{i, (r_0, r_1,b)} +\sum_{j=1}^{o}\sum_{j_1=1}^{l}\kappa^{i,j,j_1, \omega,1}_{(r_0,r_1,b)}=0, ~\textrm{for}~ (r_0, r_1,b) \in B_i \}_{2\leq \omega \leq N_{msg}}.
\end{equation}where $\kappa^{i,j,j_1, \omega,1}_{(r_0,r_1)}=t^{12}_{ij,j_1} \texttt{S}^{(j_1,\omega,1)}_{v+j,(r_0, r_1)}$. \cref{eq5} represents a linear system with $|B_i|\cdot (N_{msg}-1)$ equations and $N_{bits}\cdot l\cdot o$ unknowns over $\mathbb{F}_2$. However, the attacker does not know either $I$ or $B_i$. For the attacker to gain knowledge of $I$ and $B_i$ for $i 
\in I$, and recover $T^{12}
_{i,j}$ for $i \in I, j \in [o],$ the attacker may try to solve $N_{bits} \cdot v \cdot l^2$ linear systems separately, i.e. one for $i \in [v]$ and  $(r_0,r_1, b) \in [l]\times [l] 
\times [N_{bits}
]$, 

\begin{equation}
\label{eq6}
    \{\texttt{S}^{(\omega,1)}_{i, (r_0, r_1,b)} +\sum_{j=1}^{o}\sum_{j_1=1}^{l}\kappa^{i,j,j_1, \omega,1}_{(r_0,r_1,b)}=0 \}_{2\leq \omega \leq N_{msg}}, 
\end{equation} where each has $N_{msg}-1$ equations and $N_{bits} \cdot l \cdot o$ unknowns over $\mathbb{F}_2$. If $N_{msg}> N_{bits} \cdot l \cdot o+1$, then the linear systems for $(r_0,r_1, b)\in B_i, i \in I$ will have a  unique solution, while the other linear systems are expected to have no solution. \cref{scheme:recover_from_partial_bits} details the recovery strategy by the attacker and exploits the fact that  $\mathbb{F}_{16} \cong \mathbb{F}_2[x]\big /\langle x^4 +x+1\rangle$ for SNOVA.

Therefore, when $N_{msg}> N_{bits} \cdot lo+1$, \cref{scheme:recover_from_partial_bits} will output \texttt{dic} such that $\texttt{dic}[(i,r_0, r_1,b)]=[T^{12}_{i1},\ldots, T^{12}_{io}]$ for $i\in I, (r_0, r_1,b) \in B_i$ and $\texttt{dic}[(i,r_0, r_1,b)]=\texttt{None}$ otherwise.

\subsection{ How can the attacker  recover  $T^{12}_{i,j}$ for a fixed $i \in [v]\setminus I, j \in [o]$?} For $1 \leq \omega \leq N_{msg}$, we have

\begin{align*}
\texttt{S}^{(\omega)}_i  + \sum_{j=1}^{o}T^{12}_{ij}\texttt{S}^{(\omega)}_{v+j} =\texttt{S}^{(\omega)}_i+ \sum_{j=1}^{o}\sum_{j_1=1}^{l}t^{12}_{ij,j_1}S^{j_1-1}\texttt{S}^{(\omega)}_{v+j}&=\texttt{V}^{(\omega)}_{i}, i\in [v] \setminus I, \\
\end{align*}Note that the previous equations can always be arranged as $$\mathbf{S}_{i}+\sum_{j=1}^{o}\sum_{j_1=1}^{l}t^{12}_{ij,j_1}\mathbf{S}_{j_1,j}=\mathbf{V}_{i}$$ with $\mathbf{S}_{i}=\begin{bmatrix}
\texttt{S}_i^{(1)}\\
\vdots\\
\texttt{S}_i^{(N_{msg})}\\
\end{bmatrix}^t$, $\mathbf{S}_{j_1,j}=\begin{bmatrix}
S^{j_1-1}\texttt{S}^{(1)}_{v+j}\\
\vdots\\
S^{j_1-1}\texttt{S}^{(N_{msg})}_{v+j}\\
\end{bmatrix}^t$,   $\mathbf{V}_{i}=\begin{bmatrix}
\texttt{V}^{(1)}_{i}\\
\vdots\\
\texttt{V}^{(N_{msg})}_{i}\\
\end{bmatrix}^t \in \mathcal{R}^{1 \times N_{msg}}.$  

\bigskip

 This indeed induces an instance of the MinRank problem \cite{BUSS1999572} over $\mathbb{F}_q$. Note that by  setting $\mathbf{M}=(\mathbf{S}_{i},\mathbf{S}_{1,1}, \ldots, \mathbf{S}_{l,o} ) \in (\mathbb{F}_q^{l \times (N_{msg} \cdot l)})^{l\cdot o +1}$, there exists a $(t_{i1,1},$ $\ldots, t_{io,l}) \in \mathbb{F}_q^{ol}$ and a matrix $ \mathfrak{M} \in \mathbb{F}_{q}^{l \times (N_{msg}-1)\cdot l}$ such that 

$$(\mathbf{S}_{i}+\sum_{j=1}^{o}\sum_{j_1=1}^{l}t^{12}_{ij,j_1}\mathbf{S}_{j_1,j})\begin{bmatrix}
    \mathfrak{I}\\
    -\mathfrak{M}
\end{bmatrix}=0$$ where $\mathfrak{I} \in \mathbb{F}_{q}^{(N_{msg}-1)\cdot l \times (N_{msg}-1)\cdot l} $ is a non-singular matrix.

\begin{algorithm}[htp]  
\caption{Partially Recovers $T^{12}$ from Fixed Bits}
\small
\label{scheme:recover_from_partial_bits}
\begin{algorithmic}[1]
\STATE \textbf{function} $\texttt{recover\_partial\_T\_F2}(v,o, l,[\texttt{S}^{(1)},\ldots, \texttt{S}^{(N_{msg})}])$
\STATE $S \gets \texttt{getS}(l)$
\STATE $\texttt{dic} \gets \{\}$
\FOR{$i \gets 1$ \textbf{to} $v$}
    \FOR{$r_0 \gets 1$ \textbf{to} $l$}
        \FOR{$r_1 \gets 1$ \textbf{to} $l$}
            \FOR{$b \gets 1$ \textbf{to} 4}
                \STATE $\mathbf{A} \gets \mathbb{F}_2^{(N_{msg}-1) \times 4\cdot ol}$
                \STATE $\mathbf{Y} \gets \mathbb{F}_2^{(N_{msg}-1) \times 1}$
                \FOR{$\omega \gets 2$ \textbf{to} $N_{msg}$}
                    \STATE $\texttt{S}^{(\omega,1)}_i \gets \texttt{S}^{(1)}_i - \texttt{S}^{(\omega)}_i$
                    \FOR{$j \gets 1$ \textbf{to} $o$}
                        \FOR{$j_1 \gets 1$ \textbf{to} $l$}
                            \STATE $\texttt{S}^{(j_1,k,1)}_{v+j} \gets S^{j_1-1}(\texttt{S}^{(\omega)}_{v+j} - \texttt{S}^{(1)}_{v+j})$
                            \IF{$b=1$}
                                \STATE $\mathbf{A}_{(\omega-1),j\cdot l+4\cdot j_1+1} \gets \texttt{S}^{(j_1,k,1)}_{v+j,(r_0,r_1, 1)}$
                                \STATE $\mathbf{A}_{(\omega-1),j\cdot l+4\cdot j_1+2} \gets \texttt{S}^{(j_1,k,1)}_{v+j,(r_0,r_1, 4)}$
                                \STATE $\mathbf{A}_{(\omega-1),j\cdot l+4\cdot j_1+3} \gets \texttt{S}^{(j_1,k,1)}_{v+j,(r_0,r_1, 3)}$
                                \STATE $\mathbf{A}_{(\omega-1),j\cdot l+4\cdot j_1+4} \gets \texttt{S}^{(j_1,k,1)}_{v+j,(r_0,r_1, 2)}$
                            \ELSIF{$b=2$}
                                \STATE $\mathbf{A}_{(\omega-1),j\cdot l+4\cdot j_1+1} \gets \texttt{S}^{(j_1,k,1)}_{v+j,(r_0,r_1, 2)}$
                                \STATE $\mathbf{A}_{(\omega-1),j\cdot l+4\cdot j_1+2} \gets \texttt{S}^{(j_1,k,1)}_{v+j,(r_0,r_1, 4)} + \texttt{S}^{(j_1,k,1)}_{v+j,(r_0,r_1, 1)}$
                                \STATE $\mathbf{A}_{(\omega-1),j\cdot l+4\cdot j_1+3} \gets \texttt{S}^{(j_1,k,1)}_{v+j,(r_0,r_1, 4)} + \texttt{S}^{(j_1,k,1)}_{v+j,(r_0,r_1, 3)}$
                                \STATE $\mathbf{A}_{(\omega-1),j\cdot l+4\cdot j_1+4} \gets \texttt{S}^{(j_1,k,1)}_{v+j,(r_0,r_1, 3)} + \texttt{S}^{(j_1,k,1)}_{v+j,(r_0,r_1, 2)}$
                            \ELSIF{$b=3$}
                                \STATE $\mathbf{A}_{(\omega-1),j\cdot l+4\cdot j_1+1} \gets \texttt{S}^{(j_1,k,1)}_{v+j,(r_0,r_1, 3)}$
                                \STATE $\mathbf{A}_{(\omega-1),j\cdot l+4\cdot j_1+2} \gets \texttt{S}^{(j_1,k,1)}_{v+j,(r_0,r_1, 2)}$
                                \STATE $\mathbf{A}_{(\omega-1),j\cdot l+4\cdot j_1+3} \gets \texttt{S}^{(j_1,k,1)}_{v+j,(r_0,r_1, 1)} + \texttt{S}^{(j_1,k,1)}_{v+j,(r_0,r_1, 4)}$
                                \STATE $\mathbf{A}_{(\omega-1),j\cdot l+4\cdot j_1+4} \gets \texttt{S}^{(j_1,k,1)}_{v+j,(r_0,r_1, 4)} + \texttt{S}^{(j_1,k,1)}_{v+j,(r_0,r_1, 3)}$
                            \ELSE
                                \STATE $\mathbf{A}_{(\omega-1),j\cdot l+4\cdot j_1+1} \gets \texttt{S}^{(j_1,k,1)}_{v+j,(r_0,r_1, 4)}$
                                \STATE $\mathbf{A}_{(\omega-1),j\cdot l+4\cdot j_1+2} \gets \texttt{S}^{(j_1,k,1)}_{v+j,(r_0,r_1, 3)}$
                                \STATE $\mathbf{A}_{(\omega-1),j\cdot l+4\cdot j_1+3} \gets \texttt{S}^{(j_1,k,1)}_{v+j,(r_0,r_1, 2)}$
                                \STATE $\mathbf{A}_{(\omega-1),j\cdot l+4\cdot j_1+4} \gets \texttt{S}^{(j_1,k,1)}_{v+j,(r_0,r_1, 4)} + \texttt{S}^{(j_1,k,1)}_{v+j,(r_0,r_1, 1)}$
                            \ENDIF
                        \ENDFOR
                    \ENDFOR
                    \STATE $\mathbf{Y}_{(\omega-1),0} \gets \texttt{S}^{(\omega,1)}_{i,(r_0,r_1, b)}$
                \ENDFOR
                \STATE $ \texttt{X}, \texttt{output} \gets \texttt{Gauss}(\mathbf{A}, \mathbf{Y})$
                \IF{$\texttt{output}$}
                    \STATE $[T^{12}_{i1},\ldots, T^{12}_{io}] \gets \texttt{get\_elements\_in\_FqS\_from\_bits}(\texttt{X}, l)$
                    \STATE $\texttt{dic}[(i, r_0, r_1, b)] \gets [T^{12}_{i1}, \ldots, T^{12}_{io}]$
                \ELSE
                    \STATE $\texttt{dic}[(i, r_0, r_1, b)] \gets \texttt{None}$
                \ENDIF
            \ENDFOR
        \ENDFOR
    \ENDFOR
\ENDFOR
\RETURN $\texttt{dic}$
\STATE \textbf{end function} 
\end{algorithmic}
\end{algorithm}


\subsection{Discussion of previous scenarios}
We remark that the scenarios described in  \cref{partial_recovery_field,partial_recovery_bits} are particular cases of related randomness attacks \cite{cryptoeprint:2014/337} because what the attacker accomplishes by injecting a permanent fault is force the signature algorithm to reuse the same sub-bitstrings within the $N_{bits}vl^2$ bitstring that represent the vinegar values $\texttt{V}_1,\texttt{V}_2,\ldots, \texttt{V}_v$, even though these fixed sub-bitstrings are unknown to the attacker each time a valid signature is generated for a random message. Therefore, partially recovering the private linear map represented by $T$ is always possible by using the techniques presented in \cref{partial_recovery_field,partial_recovery_bits}, as long as the attacker finds any other means of fixing sub-bitstrings within the $N_{bits}vl^2$ bitstring and collects enough valid signatures generated using these fixed values. 

Moreover, if the attacker might distinguish a bit at a fixed position of each $\texttt{V}_i$ during signature generation, then the attacker could leverage  \cref{scheme:recover_from_partial_bits} to recover $T$ after collecting a sufficient number of signatures. Indeed, define $\mathcal{J}:=[v]\times [l]^2 \times [N_{bits}]$,  $\mathcal{G}$ and $\mathcal{O}_{\mathcal{I}}$ as shown by \cref{special_oracle}. Suppose that the attacker is given access to the set $\mathcal{I} \gets \mathcal{G}()$ and the oracle   $\mathcal{O}_{\mathcal{I}}$.



\begin{algorithm}[htp]  
\caption{defines functions $\mathcal{G}$ and $\mathcal{O}_{\mathcal{I}}$.}
\small
\label{special_oracle}
\begin{multicols}{2}

\begin{algorithmic}[1]
\small
\STATE \textbf{function} $\mathcal{G}()$
\STATE $\mathcal{I} \gets \emptyset$
\FOR{$(i\gets 1, i\leq v, i\gets i+1)$}
\STATE $(r_0,r_1,b) \xleftarrow{\$}[l]\times[l]\times [N_{bits}]$
\STATE $\mathcal{I} \gets \mathcal{I} \cup \{(i,r_0,r_1,b)\} $
\ENDFOR
\RETURN $\mathcal{I}$
\STATE \textbf{end function} 

\end{algorithmic}

\columnbreak

\begin{algorithmic}[1]
\small
\STATE \textbf{function} $\mathcal{O}_{\mathcal{I}}(\iota\in \mathcal{J}, b \in \mathbb{F}_2 )$
   
   \IF{$\iota \in  \mathcal{J}\setminus \mathcal{I}$}
   \STATE $c \xleftarrow{\$} \mathbb{F}_2$
   \RETURN $c$
   \ENDIF

   \STATE Let $\texttt{V}_{\iota}$ be the random bit chosen by the line \ref{compvin1} of 
\cref{scheme:sign} in the most recent call.

   \IF{$b = \texttt{V}_{\iota}$}
   \RETURN $1$
   \ENDIF
   
   \RETURN $0$
\STATE \textbf{end function} 
\end{algorithmic}
\end{multicols}
\end{algorithm}

This adversary can leverage his knowledge of $\mathcal{I}$,  his access to $\mathcal{O}_{\mathcal{I}}$ and  \cref{scheme:recover_from_partial_bits} to fully recover the private linear transformation $\mathcal{T}$ as follows.

\begin{enumerate}

    \item The attacker sets $\mathcal{S}=[]$, creates the lists $\texttt{L}_{\iota}=[]$ and sets $b_{\iota} \xleftarrow{\$} \mathbb{F}_2$ for all $\iota \in \mathcal{I}$.

    \item  The attacker calls  \cref{scheme:sign} for the random message $\texttt{M}^{(j)}$, which outputs $(\texttt{S}^{(j)}, \texttt{salt}^{(j)})$, and then updates $\mathcal{S}. \texttt{append}((\texttt{S}^{(j)}, \texttt{salt}^{(j)})).$

    Additionally, the attacker updates its lists $\texttt{L}_{\iota}$ for all $\iota \in \mathcal{I}$ as follows. 
    
     \begin{enumerate}
   \item \label{criticalStep}For each $\iota \in \mathcal{I}$, $\texttt{L}_{\iota}.\texttt{append}(\mathcal{O}_{\mathcal{I}}(\iota, b_{\iota}))$. 
   
    \end{enumerate}

    \item After collecting sufficient signatures, $N_{msg}$, the attacker stops. In particular, once $\sum_{i=1}^{N_{msg}}\texttt{L}_{\iota}[i]> N_{bits}\cdot l \cdot o +1$ for all $\iota \in \mathcal{I}$, it will stop. 

    \item The attacker then uses the collected signatures and calls  \cref{scheme:recover_from_partial_bits} $|\mathcal{I}|$ times to recover the matrix $T$.

     \begin{enumerate}
         \item For each $\iota=(i,r_0,r_1,b)\in \mathcal{I}$, the attacker creates $\mathcal{S}_{\iota} =[\mathcal{S}[j] \textrm{ for }  j\in [N_{msg}] \textrm{ if } \texttt{L}_{\iota}[j]=1 ]$ and calls  \cref{scheme:recover_from_partial_bits} with parameters $v,o,l$ and $\mathcal{S}_{\iota}$. From  \cref{partial_recovery_bits}, it follows each call of \cref{scheme:recover_from_partial_bits} with parameters $v,o,l$ and $\mathcal{S}_{\iota}$ will return $\texttt{dic}[\iota]=[T^{12}_{i1},\ldots, T^{12}_{io}].$

     \end{enumerate}

\end{enumerate}

We remark that the previous example scenario is yet another case of related randomness attacks, since the attacker at step \ref{criticalStep} marks what signatures share the bit $b_{\iota}$ in $\texttt{V}_{\iota}$. However, we stress that we do not know how to instantiate this oracle $\mathcal{O}_{\mathcal{I}}$ in a real scenario effectively, and therefore this question remains open.


\subsection{Fault-assisted reconciliation attack}
\label{sec:faultrecattack}

As seen in \cref{recon_attack_sec}, for the \textit{reconciliation} attack, the attacker must find a specific solution $\mathbf{u}_0 \in \mathbb{F}_q^{ln}$ from among many solutions to the quadratic system of the form
\begin{equation}
    \label{rec_fault}
\mathbf{u}_0^t(\Lambda_{S^{i}}P_k\Lambda_{S^{j}})\mathbf{u}_0=0 \in \mathbb{F}_q,
\end{equation} for~$k \in [m],i,j \in \{0,1,\ldots, l-1\}$. Furthermore, for any valid signature $(\texttt{S}, \texttt{salt})$, it holds $\texttt{S} =T^{-1}\texttt{V}$, with $\texttt{V}=(\texttt{V}_1,\ldots,\texttt{V}_v,\texttt{O}_1, \ldots, \texttt{O}_o)^t$ and $T^{-1}=T$. Consequently, for any $\beta \in [l]$, we have 

$$\texttt{S}_{:\beta}= \begin{bmatrix}
\texttt{S}_{1,:\beta} \\
\vdots\\
\texttt{S}_{v,:\beta}\\
\vdots\\
\texttt{S}_{n,:\beta}\\
\end{bmatrix}=\begin{bmatrix}
1 & 0 & \ldots & 0&T^{12}_{1,1}&\ldots &T^{12}_{1,o} \\
\vdots & \vdots & \ddots & \vdots&\vdots &\ddots &\vdots \\
0 & 0 & \ldots & 1 & T^{12}_{v,1}&\ldots &T^{12}_{v,o} \\
\vdots & \vdots & \ddots & \vdots&\vdots &\ddots &\vdots \\
0 & 0 & \ldots & \ldots &0 &\ldots&1 \\
\end{bmatrix}\begin{bmatrix}
\texttt{V}_{1,:\beta}\\
\vdots\\
\texttt{V}_{v,:\beta}\\
\texttt{O}_{1,:\beta}\\
\vdots\\
\texttt{O}_{o,:\beta}\\
\end{bmatrix}=\begin{bmatrix}
\texttt{V}_{1,:\beta} - \sum_{j=1}^{o}T^{12}_{1j}\texttt{O}_{j,:\beta}\\
\vdots\\
\texttt{V}_{v,:\beta}- \sum_{j=1}^{o}T^{12}_{vj}\texttt{O}_{j,:\beta}\\
\texttt{O}_{1,:\beta}\\
\vdots\\
\texttt{O}_{o,:\beta}\\
\end{bmatrix}$$ where $\texttt{S}_{:\beta}$ denotes the $\beta$-th column of $\texttt{S}$.  If  an attacker knows $\texttt{V}_{1,:\beta}, \ldots, \texttt{V}_{v,:\beta}$, then the attacker can set $$\mathbf{u}_0=\begin{bmatrix}
\texttt{S}_{1,:\beta} -\texttt{V}_{1,:\beta}\\
\vdots\\
\texttt{S}_{v,:\beta}-\texttt{V}_{v,:\beta}\\
\vdots\\
\texttt{S}_{n,:\beta}\\
\end{bmatrix},$$ which will satisfy \cref{rec_fault}.  Therefore, the main task of the attacker is to find $\texttt{V}_{1,:\beta}, \ldots, \texttt{V}_{v,:\beta}$ for a valid signature $(\texttt{S}, \texttt{salt})$ and some $\beta \in [l]$.

\subsubsection{Attack strategy} 
\label{sec:Fault_many}
\begin{enumerate}
    \item \label{faultmany} The attacker introduces transient faults affecting line \ref{compvin1} of \cref{scheme:sign}, specifically targeting $\texttt{V}_{i,j\beta}$ for all $i \in [v],j \in [l]$ and $\beta \in \mathcal{C} \subseteq [l]$ independently. In particular, for the variable $\texttt{V}_i$, there is a fixed non-empty subset $J_i \subseteq [l] \times \mathcal{C}$ such that $\bar{\texttt{V}}_{i,(r_0,r_1)} = \omega$ for $(r_0,r_1) \in J_i$, where $\omega \in \mathbb{F}_{q}$ is some unknown and fixed value.
    
   \item \label{signature:col} The attacker then calls \cref{scheme:sign} with a randomly chosen message $\texttt{M} \in \mathcal{R}^{m}$ and receives the signature $(\texttt{S}, \texttt{salt})$.

   \item If Step \ref{signature:col} succeeds, then call Algorithm \ref{fault:esp} with parameters $v,o,l,S, \mathcal{C},\Gamma_{\beta}$ for $\beta \in \mathcal{C}$\footnote{If $\mathcal{C}$ is unknown, the attacker can always set $\mathcal{C}=[l]$. }, where  $\Gamma_{\beta} \subseteq [lv]$. Furthermore, \( F_{\gamma} \) denotes  the set of all subsets of \( \gamma \) integers that can be selected from \( [lv] \) and \( A^c = [lv] \setminus A \) with \( A\in F_{\gamma} \).  
\end{enumerate}

We remark that by choosing proper $\Gamma_{\beta}$'s, the attacker can ensure that the quadratic systems to be solved at the line \ref{quadratic_system} of Algorithm \ref{fault:esp} have $ol^2$ equations and $lv-\gamma < ol^2$ unknowns. Thus, they are expected to have either no solution or very few solutions.

Let $\texttt{V}=(\texttt{V}^t_1, \ldots, \texttt{V}^t_v)^t \in \mathbb{F}_q^{vl^2}$.   If after carrying out Steps \ref{faultmany} and \ref{signature:col}, there exists $\beta \in \mathcal{C}$ such that $\texttt{V}_{i\beta}=\omega$, for $i \in A$, with $A \in F_{\gamma}$ and $\gamma \in \Gamma_{\beta}$, then Algorithm \ref{fault:esp} will find an $\texttt{U}$ satisfying \cref{rec_fault} and an secret space $\mathcal{O}$. Otherwise, the attacker may start the attack again. In \cref{runtime_algorithm_fault}, we analyse Algorithm \ref{fault:esp}'s runtime complexity.

\begin{algorithm}[htp]  
\caption{attempts to find $\mathcal{O}$ after having run the attack strategy.}
\label{fault:esp}
\begin{algorithmic}[1]

\STATE \textbf{function} $\texttt{fault\_assisted\_reconcilation\_attack}(v,o, l, \texttt{S}, \mathcal{C}, \Gamma_{\beta} \textrm{ for } \beta \in \mathcal{C} )$

\FOR{$\beta \in \mathcal{C}$ }

\FOR{ $\gamma \in \Gamma_{\beta}$}

\FOR{ $A \in F_{\gamma}$ }

\FOR{$\omega \in \mathbb{F}_{q}$}
\STATE Set $\Omega \gets (x_1,\ldots, x_{lv}, 0, \ldots, 0)^t \in \mathbb{F}_q^{ln}$
\STATE Set $\Omega_i \gets w $ for $i \in A$ 
 
\STATE $\texttt{X} \gets \texttt{S}_{:\beta} -\Omega$
\STATE \label{quadratic_system}Attempt to solve the quadratic system 

$$\texttt{X}^t(\Lambda_{S^{i}}P_k\Lambda_{S^{j}})\texttt{X}=0 \in \mathbb{F}_q,$$
 for~$k \in [m],i,j \in \{0,1,\ldots, l-1\}$. This system has $ml^2$ equations and $lv-\gamma$ unknowns, namely $x_i$ for $i \in A^c$ 
 
 \IF{$x_i$ for $i \in A^c$ are found}
  \STATE Set $\texttt{U}$ as the solution.
  \STATE  Recover $\mathcal{O}$ from the linearly independent set $\{\Lambda_{S^{j}}\texttt{U}: 0 \leq j\leq l-1 \}$
   \RETURN $\mathcal{O}$
\ENDIF

\ENDFOR 
\ENDFOR 

\ENDFOR 
 \ENDFOR
  \RETURN $\perp$
\end{algorithmic}
\end{algorithm}

\paragraph{Success Probability of our Attack Strategy.}For $i \in [vl], j\in \mathcal{C}$, let \( X_{ij} \in \{0,1\} \) be a Bernoulli random variable that indicates whether \( \texttt{V}_{ij} \) is fixed to \( \omega \) due to the corresponding transient fault. Let \( Pr(X_{ij}) = p_{ij} \) be the probability that \( X_{ij} = 1 \), i.e., that \( \texttt{V}_{ij} \) is fixed to \( \omega \).
Define \( Y_{\beta} = \sum_{i=1}^{lv} X_{i\beta} \). The probability of obtaining \( 0\leq \gamma \leq lv \) successful fixes out of a total of \( lv \) in the $\beta$-th column of $\texttt{V}$ can be expressed as

\[
\Pr(Y_{\beta} = \gamma) = \sum\limits_{A \in F_{\gamma}} \prod\limits_{i \in A} p_{i\beta} \prod\limits_{j \in A^c} (1 - p_{j\beta}).
\] 

Let \( \rho_{\beta} \) be the probability there exists $\gamma \in \Gamma_{\beta}$ such that $Y_{\beta}=\gamma$. Therefore,

\[ \rho_{\beta} = \sum_{\gamma \in \Gamma_{\beta}}\Pr(Y_{\beta}=\gamma ) . \] 

Let \( \rho \) denote \textit{the success probability of Algorithm \ref{fault:esp}}, meaning there exists $\beta \in \mathcal{C}$ such that $\texttt{V}_{i\beta}=\omega$, for $i \in A$, with $A \in F_{\gamma}$ and $\gamma \in \Gamma_{\beta}$. Therefore, \[ \rho = \Pr(Y_{\beta} = \gamma \text{ for some } \beta \in \mathcal{C} \text{ with } A \in F_{\gamma} \text{ and } \gamma \in \Gamma_{\beta}  ) =\max\{p_{\beta}: \beta \in \mathcal{C}\}.\]

Overall, a run of the attack strategy is successful with probability \( \rho(1 - \delta) \), where \( \delta \) is the failure probability of Step \ref{signature:col} of the attack strategy (i.e., when Algorithm \ref{scheme:sign} fails to return a signature after one iteration). Furthermore, if the \( p_{ij} \)'s remain constant for each run of the attack strategy, the attacker is expected to execute the attack strategy \( 1/\rho(1 - \delta) \) times.

We note that Step \ref{faultmany} of this attack strategy can potentially be implemented by inducing a single transient fault, similar to the one found in the C implementation of Keccak used to generate the vinegar variables for MAYO \cite{jendral2024p}. For SNOVA, such a transient fault would target \( \texttt{V}_{i\beta} \) for $i\in[lv], \beta\in [l]$. Therefore, the attacker would need to run the attack strategy \( 1/(1 - \delta) \) times if $p_{i\beta}=1$ for all $i$ and $\beta$. 
However, if the attacker only can ensure \( \epsilon \leq p_{i\beta} \leq  1 \) with $\epsilon \geq 0$, they can choose some  $\Gamma_{\epsilon}$  for all $\beta\in [l]$, and hence

\[
(1-\delta)\sum\limits_{\gamma \in \Gamma_{\epsilon}}  {lv \choose \gamma}\epsilon^\gamma (1-\epsilon)^{lv-\gamma} \leq (1-\delta)\max\{\rho_{\beta}= \sum\limits_{\gamma \in \Gamma_{\epsilon}} \Pr(Y_{\beta}=\gamma): \beta \in \mathcal{C}\} \leq (1-\delta).
\]

Moreover, the attacker might improve the attack strategy's success probability by inducing transient faults that specifically target \( \texttt{V}_{i\beta} \) for all \( i \in [lv] \), and \( \beta \in \mathcal{C} \) with $|\mathcal{C}|=1$. In such a case, \( \delta \) is expected to be very low. Therefore, if the attacker only can ensure \( \epsilon \leq p_{i\beta} \leq  1 \) for a $\epsilon\geq 0$, they might set a proper $\Gamma_{\epsilon}$ such that $\sum\limits_{\gamma \in \Gamma_{\beta}}  {lv \choose \gamma}\epsilon^\gamma (1-\epsilon)^{lv-\gamma} \approx 1$, and expect to run the attack strategy\footnote{The runtime of Algorithm \ref{fault:esp} depends on $\Gamma_{\epsilon}$.} only once. We analyze various scenarios in our simulations in \cref{sim_fault_rec_attack}.

\subsection{Alternative Versions of SNOVA}

The SNOVA team recently released a preprint \cite{snova_modified} that proposes two new versions of SNOVA to counteract Buellen's attack~\cite{Beu24}.

The first alternative version of SNOVA is choose random matrices $A_{k,\alpha},B_{k,\alpha} \in \mathcal{R}$ and $Q_{k,\alpha1}, Q_{k,\alpha 2} \in \mathbb{F}_q[S]$, for $k \in [o]$ and $\alpha \in [l^2]$, and define the $k$-th coordinate of the public map $\cP(U)$ as
\begin{equation*}
 \mathcal{P}_k ( U_1, \ldots, U_n) =  \sum_{\alpha=1}^{l^2} \sum_{i=1}^{n} \sum_{j=1}^{n} A_{k,\alpha}\cdot U^t_{i}(Q_{k,\alpha1}P_{k, i,j}Q_{k,\alpha 2})U_{j}\cdot B_{k,\alpha}.
\end{equation*}

The second alternative version of SNOVA defines the $k$-th coordinate of the public map $\cP(U)$ as follows

\begin{equation*}
 \mathcal{P}_k (U)=  \sum_{\alpha=1}^{l^4} \sum_{i=1}^{n} \sum_{j=1}^{n} A_{\alpha}\cdot U^t_{i}(Q_{\alpha1}P_{k, i,j}Q_{\alpha 2})U_{j}\cdot B_{\alpha}, 
\end{equation*} where the matrices $A_{\alpha},B_{\alpha} \in \mathcal{R}$, and $Q_{\alpha1}, Q_{\alpha 2} \in \mathbb{F}_q[S]$, for  $\alpha \in [l^4]$, are determined by fixed matrices  $\tilde{E}_{i,j} \in \mathbb{F}_q^{l^2\times l^2}$, for $i,j \in [l]$, specified in \cite{snova_modified}.

We remark that either alternative version is \emph{still vulnerable to our fault attacks} described in \cref{partial_recovery_field,partial_recovery_bits} since we exploit the related randomness present in the vinegar variables $\texttt{V}^{(\omega)}=(\texttt{V}_1^{(\omega)},\ldots, \texttt{V}_n^{(\omega)})^t$ when a permanent fault has been established and the relation  $\texttt{S}^{(\omega)} =T^{-1}\texttt{V}^{(\omega)}$. Additionally, the proposed alternatives do not affect the reconciliation attack.

\section{Experiments of our fault attacks}
\label{sec:simulations}
To validate our claims, we conducted simulations of our fault attack and presented a step-by-step procedure for 
our tests along with their results. We begin by explaining the conceptual framework of the attack. We implement the SNOVA signature scheme in SAGE, adhering to its specification document~\cite{SNOVA}.  This implementation serves as the foundation for our analysis. Additionally, we utilized the latest version 
of the SNOVA code provided by the SNOVA team\footnote{\url{https://github.com/PQCLAB-SNOVA/SNOVA} using commit 
3d7e8c7cebdd57293d74dc6c2608656697b99597.} to generate the signatures. In these simulations, we introduced faults 
by fixing specific values in the vinegar variables, thereby mimicking the fault injection process described in 
our attack model in \cref{sec:adversary_model}.



\subsection{Evaluating the fault attack from \cref{partial_recovery_field,partial_recovery_bits} }
We evaluate our attack by considering two scenarios.

In \textit{Scenario I}, we replace line~\ref{compvin1} of Algorithm~\ref{scheme:sign} with the function described in Algorithm~\ref{fault:simf16}. This function takes a set of SNOVA parameters, a binary string $\texttt{x}$ of size $l^2v$, and a list $\texttt{L}$ of the same size containing random field elements from $\mathbb{F}_{16}$. The binary string $\texttt{x}$ directs Algorithm~\ref{fault:simf16} regarding which elements are to be drawn from $\texttt{L}$ and which are to be generated randomly. This approach ensures that each time the signature algorithm is executed, the same field elements in each $\texttt{V}_i$ remain fixed.

\begin{algorithm}[htp]  
\caption{simulates a fault by fixing $\mathbb{F}_{16}$ elements in the vinegar variables.}
\label{fault:simf16}
\begin{algorithmic}[1]

\STATE \textbf{function} $\texttt{assign\_values\_to\_vinegar\_variables\_fault\_F16}(v,o, l, \texttt{x}, \texttt{L})$

\STATE  $\texttt{V} \gets []$
\FOR{ ($i\gets 1, i\leq v, i \gets  i+1$)}
\STATE $\texttt{V}_i \gets [0]^{l\times l}$
\FOR{($r_0\gets 1, r_0\leq l, r_0 \gets  r_0+1)$}
\FOR{($r_1\gets 1, r_1\leq l, r_1 \gets  r_1+1)$}
\IF{($\texttt{x}_{i\cdot l^2+r_0 \cdot l+r_1} =1$)}

\STATE $\texttt{V}_i[r_0,r_1]\gets \texttt{L}_{i \cdot l^2+r_0 \cdot l+r_1}$
\ELSE
\STATE $\texttt{V}_i[r_0,r_1]\xleftarrow{\$} \mathbb{F}_{16}$ 

\ENDIF
\ENDFOR 

\ENDFOR   
\STATE $\texttt{V}.\texttt{append}(\texttt{V}_i)$
 \ENDFOR
 \RETURN $\texttt{V}$
\end{algorithmic}
\end{algorithm}

In \textit{Scenario II}, we replace line~\ref{compvin1} of Algorithm~\ref{scheme:sign} with the function represented by Algorithm~\ref{fault:simf2}. This function takes a SNOVA parameter set, a binary string $\texttt{x}$ of size $4l^2v$, and a list $\texttt{L}$ of size $4l^2v$ containing random field elements from $\mathbb{F}_{2}$. Similar to the previous scenario, Algorithm~\ref{fault:simf2} guarantees that the same bits in the binary representation of each $\texttt{V}_i$ are fixed each time Algorithm~\ref{scheme:sign} is executed.

\begin{algorithm}[htp]  
\caption{simulates a fault by fixing $\mathbb{F}_{2}$ elements in the vinegar variables.}
\label{fault:simf2}
\begin{algorithmic}[1]

\STATE \textbf{function} $\texttt{assign\_values\_to\_vinegar\_variables\_fault\_F2}(v,o, l, \texttt{x}, \texttt{L})$

\STATE  $\texttt{V} \gets []$
\FOR{ ($i\gets 1, i\leq v, i \gets  i+1$)}
\STATE $\texttt{V}_i \gets [0]^{l\times l}$
\FOR{($r_0\gets 1, r_0\leq l, r_0 \gets  r_0+1)$}
\FOR{($r_1\gets 1, r_1\leq l, r_1 \gets  r_1+1)$}
\STATE $\texttt{e} \gets [0]^4$
\FOR{($r_2\gets 1, r_2\leq 4, r_2 \gets  r_2+1)$}

\IF{($\texttt{x}_{i\cdot l^2+r_0\cdot l+4\cdot r_1+r_2} =1$)}

\STATE $\texttt{e}[r_2]\gets \texttt{L}_{i\cdot l^2+r_0\cdot l+4\cdot r_1+r_2}$
\ELSE
\STATE $\texttt{e}[r_2]\xleftarrow{\$} \mathbb{F}_{2}$ 

\ENDIF

\ENDFOR 

\STATE $\texttt{V}_i[r_0,r_1] \gets \texttt{e}$
\ENDFOR  
\ENDFOR   
\STATE $\texttt{V}.\texttt{append}(\texttt{V}_i)$
 \ENDFOR
 \RETURN $\texttt{V}$
\end{algorithmic}
\end{algorithm}

In summary, our \emph{test experiments}  run the following step-by-step procedure:

\begin{enumerate}
    \item Select a SNOVA parameter set and then create a key pair $(\texttt{sk},\texttt{pk})$ by calling Algorithm \ref{scheme:keygen}.

    \item In either scenario, create the bitstring $\texttt{x}$  by performing either $l^2v$ or $4\cdot l^2v$ Bernoulli trials given a probability $0<\rho <1$. Furthermore, create $\texttt{L}$ by generating a list of size $|x|$ of random field elements (from either $\mathbb{F}_{16}$ or $\mathbb{F}_{2}$).
   \item Collect $N_{msg}$ signatures by calling the tweaked version of Algorithm~\ref{scheme:sign}. The value of $N_{msg}$ depends on the scenario we are running. For the scenario I, $N_{msg}$ is set to $o\cdot l +2$, while, for the scenario II, $N_{msg}$ is set to $4\cdot o\cdot l +2$

   \item Once $N_{msg}$ signatures are collected, then call the corresponding recovery algorithm, i.e. either Algorithm \ref{scheme:recover_from_partial} for Scenario I or Algorithm \ref{scheme:recover_from_partial_bits} for Scenario II.

   \item Compare the recovered part of $T$ with the corresponding part of the real $T$ to verify the effectiveness of our recovery algorithms.

\end{enumerate}

Our \textit{experimental results} confirm that our recovery algorithms perform as expected. Specifically, when provided with the required number of faulty signatures, they successfully recover the correct components of \( T \), as outlined in Sections~\ref{partial_recovery_field} and~\ref{partial_recovery_bits}. Table~\ref{tab:simresults} summarizes the minimum number of faulty signatures required for each SNOVA parameter set to guarantee the success of our recovery algorithms.

\begin{table}[htp]
\centering
\caption{Minimum number of signatures per SNOVA parameter set}
\label{tab:simresults}
\begin{tabular}{c|c|c|c}
\toprule
 &  & Recovery from  & Recovery from  \\ 
 Security& $(v, o, q, l, \lambda)$ & fixed field elements & fixed bits   \\
 Level& & by Algorithm \ref{scheme:recover_from_partial} & by Algorithm \ref{scheme:recover_from_partial_bits}   \\
\midrule
\multirow{3}{*}{I}   & $(37, 17, 16, 2, 128)$ & $34$  &  $138$     \\
                     & $(25, 8, 16, 3,128)$  & $26$  & $98$    \\
                     & $(24, 5, 16, 4,128)$  & $22$  & $82$      \\ 
\midrule
\multirow{3}{*}{III} & $(56, 25, 16, 2,192)$ & $52$ & $202$    \\
                     & $(49, 11, 16, 3, 192)$ & $35$& $134$      \\
                     & $(37, 8, 16, 4, 192)$  & $34$  & $130$     \\ 
\midrule
\multirow{3}{*}{V}   & $(75, 33, 16, 2, 256)$ & $68$ & $266$   \\
                     & $(66, 15, 16, 3, 256)$ & $47$& $182$    \\
                     & $(60, 10, 16, 4, 256)$ & $42$  & $162$      \\ 
\bottomrule
\end{tabular}
\end{table}

\paragraph{In addition to the SAGE implementation,} we utilize the C version of the SNOVA code to evaluate our attack by generating faulty signatures. Specifically, we incorporate Algorithm~\ref{fault:simf16} into the C code. The vinegar values are generated within the function \texttt{sign\_digest\_core\_ref} found in the file \texttt{snova\_kernel.h}. To generate these values, SNOVA uses Keccak, and in the reference implementation, the authors rely on the eXtended Keccak Code Package (XKCP)\footnote{\url{https://github.com/XKCP/XKCP}}. Listing~\ref{lst:vinegar} details the method for generating these values and demonstrates how they are assigned to the appropriate structure, a matrix variable named \textsc{X\_in\_GF16Matrix}.

\begin{lstlisting}[language=C, label=lst:vinegar, caption=Code snippet of generation of vinegar values.]
 // generate the vinegar value
        Keccak_HashInstance hashInstance;
        Keccak_HashInitialize_SHAKE256(&hashInstance);
        Keccak_HashUpdate(&hashInstance, pt_private_key_seed, 8 * seed_length_private);
        Keccak_HashUpdate(&hashInstance, digest, 8 * bytes_digest);
        Keccak_HashUpdate(&hashInstance, array_salt, 8 * bytes_salt);
        Keccak_HashUpdate(&hashInstance, &num_sign, 8);
        Keccak_HashFinal(&hashInstance, NULL);
        Keccak_HashSqueeze(&hashInstance, vinegar_in_byte, 8 * ((v_SNOVA * lsq_SNOVA + 1) >> 1));

        counter = 0;
        for (int index = 0; index < v_SNOVA; index++) {
            for (int i = 0; i < rank; ++i) {
                for (int j = 0; j < rank; ++j) {
                    set_gf16m(X_in_GF16Matrix[index], i, j,
                              ((counter & 1) ? (vinegar_in_byte[counter >> 1] >> 4) : (vinegar_in_byte[counter >> 1] & 0xF)));
                    counter++;
                }
            }
        }
\end{lstlisting}

To generate faulty signatures, we employ the \texttt{get\_F16} function, which generates random field elements in \(\mathbb{F}_{16}\) and creates a binomial random variable array, denoted as \(x\). This array plays a crucial role in determining which entries in the matrix \(\textsc{X\_in\_GF16Matrix}\) will be assigned random values versus those that will be derived from the vinegar variables. As illustrated in Listing~\ref{lst:set}, we modify the variable \texttt{X\_in\_GF16Matrix} to incorporate these random values, replacing the original values obtained from the hash function. For more details about the implementation of the function \texttt{get\_F16}, see Appendix~\ref{app:code_gf}.

\begin{lstlisting}[language=C, label=lst:set, caption=Code snippet for setting random values as Algorithm 6.]
uint8_t x[v_SNOVA*lsq_SNOVA] = {0};
uint8_t I[v_SNOVA*lsq_SNOVA] = {0};
get_F16(v_SNOVA, o_SNOVA, I, x, 0.5);
for (int index = 0; index < v_SNOVA; index++) {
            for (int i = 0; i < rank; ++i) {
                for (int j = 0; j < rank; ++j) {
                    if (x[index * lsq_SNOVA + i * l_SNOVA + j] == 1) {
                        set_gf16m(X_in_GF16Matrix[index], i, j, I[index * lsq_SNOVA+ i * l_SNOVA + j]);
                    } else {
                        set_gf16m(X_in_GF16Matrix[index], i, j,
                                  ((counter & 1) ? (vinegar_in_byte[counter >> 1] >> 4) : (vinegar_in_byte[counter >> 1]
                                      & 0xF)));
                        counter++;
                    }
                }
            }
        }
\end{lstlisting}

Now, we use the faulty `\textsc{X\_in\_GF16Matrix}' to generate the signature, where the matrix `T12' is part of the private key. The process is shown in Listing~\ref{lst:signature}. Specifically, the matrix `T12' is multiplied with parts of the `\textsc{X\_in\_GF16Matrix}`, and the results are accumulated to form the final signature.

\begin{lstlisting}[language=C, label=lst:signature, caption= Usage of \textsc{X\_in\_GF16Matrix} to generate the final signature.]
for (int index = 0; index < v_SNOVA; ++index) {
    gf16m_clone(signature_in_GF16Matrix[index], X_in_GF16Matrix[index]);
    for (int i = 0; i < o_SNOVA; ++i) {
        gf16m_mul(T12[index][i], X_in_GF16Matrix[v_SNOVA + i], gf16m_secret_temp0);
        gf16m_add(signature_in_GF16Matrix[index], gf16m_secret_temp0, signature_in_GF16Matrix[index]);
    }
}
for (int index = 0; index < o_SNOVA; ++index) {
    gf16m_clone(signature_in_GF16Matrix[v_SNOVA + index], X_in_GF16Matrix[v_SNOVA + index]);
}
\end{lstlisting}

As one expects, the results obtained from SAGE—when generating signatures and employing our recovery algorithms—are quite comparable to those produced by the C code for generating faulty signatures, followed by the execution of recovery algorithms in SAGE. The C code and SAGE scripts are publicly accessible in this repository: \url{https://github.com/gbanegas/fault\_sim\_snova}.

\subsection{Evaluating the fault-assisted reconciliation attack}
\label{sim_fault_rec_attack}

Using our SAGE implementation and the C version, we also evaluated the fault attack described in \cref{sec:faultrecattack}. 

Let $P$ be a matrix of size $v \times l \times l$, where the elements $P_{ijk}$ represent the probabilities $p_{i,jk}$ as described in \cref{sec:Fault_many}. We replace line~\ref{compvin1} of Algorithm~\ref{scheme:sign} with a function that takes a set of SNOVA parameters and the matrix $P$, randomly selects $\omega \in \mathbb{F}_q$ and returns the vinegar variables $\texttt{V}_i$, where each $\texttt{V}_{i,jk}$ is equal to  $\omega$  with probability $P_{ijk}$.  

We conducted experiments, each consisting of 100 runs of SNOVA and our algorithms. In each trial, probabilities for $P_{i,jk}$ are set, and the modified version of Algorithm~\ref{scheme:sign} is run. A ``failure'' in Step \ref{signature:col} occurs if the modified algorithm cannot compute a signature after one iteration. A success in Algorithm \ref{fault:esp} occurs if the secret space can be computed after Step \ref{signature:col} has completed successfully. Therefore, the success rate of Algorithm \ref{fault:esp} is the number of successful computations divided by the number of trials, excluding those that failed in Step \ref{signature:col}. Finally, the overall success rate of the attack strategy is the number of successes in Algorithm \ref{fault:esp} divided by the total number of trials. 

In our experiments we set $\epsilon \in \{ 1, 0.97, 0.95, 0.93\}$ and $\Gamma_{\epsilon,r}=\{\lfloor \mu+r\sigma\rfloor\ , \ldots, \lfloor \mu-r\sigma\rfloor\}$, where $\mu=lv\epsilon$, $\sigma=\sqrt{lv  \epsilon(1-\epsilon)}$ and $r\in\{1,2\}$. Table \ref{table:simresults_rec} shows our results for different assignment for $P$ and the SNOVA parameter $(37, 17, 16, 2, 128)$. \cref{fault:esp}'s runtime was computed by using \cref{compl_key_recvery_rec_fault} and the Cryptographic Estimators library \cite{cryptographicestimators}. 
\begin{table}[htp]
\centering

\caption{ Table with the results of our experiments for the SNOVA parameter $(37, 17, 16, 2, 128)$, where $C_{\beta}:=\{(i,j,\beta): i \in [v],j\in[l]\}$ for some $\beta \in [l]$.}
\resizebox{\textwidth}{!}{%
\begin{tabular}{|c|c|c|c|c|c|c|c|}
 \hline
    Assignments for $P$ &
      Failure Rate &
      \multicolumn{2}{c|}{\cref{fault:esp}}  &
      \multicolumn{2}{c|}{Attack Strategy}&\multicolumn{2}{c|}{\cref{fault:esp}} \\
      & Step \ref{signature:col}
      &
      \multicolumn{2}{c|}{Success Rate}  &
      \multicolumn{2}{c|}{Success Rate}& \multicolumn{2}{c|}{Runtime (bits)}\\
     
    & & $r=1$ &$r=2$ &$r=1$ & $r=2$ &$r=1$ & $r=2$ \\
    \hline
    $P_{\iota}=1$ for $\iota \in [v]\times[l]^2$   & $6$\% & $100$\% & $100$\% & $94$\% & $94$\% & $6$ & $7$ \\
    \hline
    $0.97\leq P_{\iota}\leq1$ for $\iota \in [v]\times[l]^2$ &$6$\% &$44$\% & $100$\% & $41$\% & $94$\% & $42$& $52$\\
    \hline
   $0.95\leq P_{\iota}\leq1$ for $\iota \in [v]\times[l]^2$  & $7$\% & $33$\% & $100$\% & $31$\% & $93$\% & 52& 60\\
    \hline
    $0.93\leq P_{\iota}\leq1$ for $\iota \in [v]\times[l]^2$  & $5$\% & $19$\% & $100$\% & $18$\% & $95$\% & 60& 67\\
    \hline
   
    $ P_{\iota}=1$ for $\iota \in C_{\beta}$   &  &  &  &  &  &  &  \\
    $P_{\iota}=1/q$ for $\iota \in [v]\times[l]^2 \setminus C_{\beta}$   & $2$\% & $100$\% & $100$\% & $98$\% & $98$\% & $5$ & $6$ \\
    \hline
     
    $0.97\leq P_{\iota} \leq 1$ for $\iota \in C_{\beta}$   &  &  &  &  &  &  &  \\
    $P_{\iota}=1/q$ for $\iota \in [v]\times[l]^2 \setminus C_{\beta}$   & $8$\% & $41$\% & $100$\% & $38$\% & $92$\% & $41$ & $51$ \\
    \hline
    
    $0.95\leq P_{\iota} \leq 1$ for $\iota \in C_{\beta}$   &  &  &  &  &  &  &  \\
    $P_{\iota}=1/q$ for $\iota \in [v]\times[l]^2 \setminus C_{\beta}$   & $5$\% & $37$\% & $100$\% & $35$\% & $95$\% & $51$ & $59$ \\
    \hline
     $0.93\leq P_{\iota} \leq 1$ for $\iota \in C_{\beta}$   &  &  &  &  &  &  &  \\
    $P_{\iota}=1/q$ for $\iota \in [v]\times[l]^2 \setminus C_{\beta}$   & $4$\% & $40$\% & $93$\% & $38$\% & $89$\% & $59$ & $66$ \\
    \hline

\end{tabular}
}
\label{table:simresults_rec}
\end{table}

\section{Countermeasure }
\label{sec:countermeasure}

In this section, we present a countermeasure for the fault attacks described in Section \ref{sec:fault}.

This countermeasure adapts a general strategy designed to defend against fault attacks targeting multivariate public key cryptosystems. This strategy was initially proposed in \cite{10.1007/978-3-642-25405-5_1} and later extended and tailored for the UOV and Rainbow schemes in \cite{kramer2019}.

Specifically, Algorithm \ref{fault:countermeasure1} implements this countermeasure for SNOVA and should be invoked by Algorithm \ref{scheme:sign} immediately after executing line \ref{compvin1}.

Algorithm \ref{fault:countermeasure1} accepts three positive integers, \(\Gamma\) and \(\Lambda\), with the condition that \(\Gamma < \Lambda\), and $\Upsilon$, as well as a tuple of finite field elements \((\alpha_1, \ldots, \alpha_{l^2v})\) of size \(l^2v\). This tuple represents the SNOVA vinegar values \([\texttt{V}_1, \texttt{V}_2, \ldots, \texttt{V}_v]\) generated at line \ref{compvin1} of Algorithm \ref{scheme:sign}. Furthermore, the function \texttt{compare}, called by Algorithm \ref{fault:countermeasure1} at line \ref{counter1:compare}, takes two tuples of size \(l^2v\): \((\alpha_1, \ldots, \alpha_{l^2v})\) and \((\beta_1, \ldots, \beta_{l^2v})\). It returns a tuple of size \(l^2v\) where the \(j\)-th entry is \(1\) if \(\alpha_j \neq \beta_j\) and \(0\) otherwise. Finally, the function $\texttt{checkColumn}$ takes $(\alpha_1, \ldots, \alpha_{l^2v}),x, \beta$, $\Upsilon$ and checks if there are at least $\Upsilon$ occurrences of $x$ in the sequence  $\texttt{V}_{1,1\beta}, \ldots, \texttt{V}_{1,l\beta}, \ldots, \texttt{V}_{v,l\beta}$.

\begin{algorithm}[htp]  
\caption{Countermeasure by checking and storing vinegar values}
\label{fault:countermeasure1}
\begin{algorithmic}[1]

\STATE \textbf{function} $\texttt{countermesure}(\Gamma, \Lambda,\Upsilon, (\alpha_1, \ldots, \alpha_{l^2v}))$

   \begingroup
    \color{blue}
\label{con_rec1}\FOR{$x \in \mathbb{F}_q$}

   \FOR{$\beta \in \{1,2,\ldots, l\}$}
   \label{check_con} \IF{$\texttt{checkColumn}((\alpha_1, \ldots, \alpha_{l^2v}),x, \beta, \Upsilon)$}
    \RETURN \texttt{fail}
    \ENDIF
   
   \ENDFOR
    
\ENDFOR\label{con_rec2}
\endgroup

\IF{$\texttt{L}$ has not been created}
  \STATE $\texttt{L}\gets []$
\ENDIF
\STATE $\texttt{count} \gets [0]^{vl^2}$
\FOR{$(i\gets 0, i<|\texttt{L}|, i\gets i+1)$}
   \STATE \label{counter1:compare} $\texttt{count} \gets 
   \texttt{count} + \texttt{compare}(\texttt{L}[i],(\alpha_1, \ldots, \alpha_{l^2v}))$
\ENDFOR
   \begingroup
    \color{blue}
\IF{$\texttt{count}[j]>\Gamma$ for some $j\in[l^2v]$}
  \RETURN \texttt{fail} \label{check_con1}
\ENDIF
\endgroup

\IF{$|\texttt{L}|=\Lambda$}
  \STATE $\texttt{L}.\texttt{removeEntryAtIndex}(0)$
\ENDIF

\STATE $\texttt{L}.\texttt{append}((\alpha_1, \ldots, \alpha_{l^2v}))$
\RETURN \texttt{success}

\end{algorithmic}
\end{algorithm}

\paragraph{Why does this countermeasure work?}

Let us assume that the countermeasure is already in place in the signing algorithm, with \(\Lambda = l \cdot o\) and \(\Gamma < \Lambda\). 

We begin by analyzing the check performed between lines \ref{con_rec1} and \ref{con_rec2}, which targets the attack strategy outlined in \cref{sec:faultrecattack}. Let $\texttt{V}=(\texttt{V}_1^t,\ldots, \texttt{V}_v^t)^t$ and let $Z_{\beta}$ be the random variable that counts the number of occurrences of $x \in \mathbb{F}_q$ in the $\beta$-th column of $\texttt{V}$. $Z_{\beta}$ follows a binomial distribution with probability $p=1/q$ in the absence of faults. Therefore, we have to take $\Upsilon$  such that $\Pr(Z_{\beta}\geq \Upsilon)$, the probability of \texttt{checkColumn} returning $\texttt{True}$ at line \ref{check_con}, is negligible. For any current SNOVA parameters, $\Upsilon=\lfloor l\cdot v \cdot p+ r \sqrt{l\cdot v \cdot p(1-p)} \rfloor$, where $r\geq 6$ is natural number.   

On the one hand, if the step described in \cref{sec:Fault_many} successfully fixes at least $\Upsilon$ entries in the $\beta$-th column of $\texttt{V}$ to $x$, the function \texttt{checkColumn} will return \texttt{True}. On the other hand, if the faults fix at most $\Upsilon - 1$ entries in any column of $\texttt{V}$ to $x$, the corresponding checks will return \texttt{False}. In this case, the attacker can proceed with the attack strategy. However, running Algorithm \ref{fault:esp} under these conditions would lead to a prohibitive runtime, as discussed in Appendix \ref{runtime_algorithm_fault}.

We now examine the countermeasure against the attack strategy described in~\cref{partial_recovery_field}.

Assume that \(|\texttt{L}| = \Lambda\), meaning the list maintained by Algorithm \ref{fault:countermeasure1} has reached its maximum allowed size, and that the \texttt{countermeasure} function is called with parameters \(\Gamma\), \(\Lambda\), and \((\alpha_1, \ldots, \alpha_{l^2v})\).

Let \(E\) denote the event where line \ref{check_con1} of Algorithm \ref{fault:countermeasure1} returns \texttt{fail} in the absence of fault injection. For this event to occur, there must exist some \(j \in [l^2v]\) such that \(\texttt{count}[j] > \Gamma\), meaning the tuple \((\texttt{L}[i][j])_{i=0}^{\Lambda-1}\) contains \(\texttt{count}[j]\) instances of \(\alpha_j\). Let \(X_j\) be the random variable that counts the number of occurrences of \(\alpha_j\) in the tuple \((\texttt{L}[i][j])_{i=0}^{\Lambda-1}\). Each \(X_j\) follows a binomial distribution with probability \(p_j = \frac{1}{|\mathbb{F}_q|}\), and the variables are mutually independent. Therefore, the probability that Algorithm \ref{fault:countermeasure1} reaches line \ref{check_con1} is

\begin{align*} 
p_{\text{fail}} &= \Pr(X_j > \Gamma \text{ for some } j \in [l^2v]) \\
&= 1 - \Pr(X_j \leq \Gamma \text{ for all } j \in [l^2v]).
\end{align*}

On the other hand, if an attacker introduces a permanent fault injection, as described in Section \ref{partial_recovery_field}, the signing algorithm will abort after producing \(\Gamma\) faulty signatures. In this case, the attacker could construct \(v^2l\) linear systems (as in Eq. (\ref{eq4})), each consisting of \(\Gamma - 1\) equations and \(l o\) unknowns. Since \(\Gamma - 1 < \Lambda = l o\), these linear systems are under-determined, meaning they have multiple solutions. Moreover, the attacker is unaware of which field elements, if any, from each vinegar variable \(\texttt{V}_i\) are fixed by the permanent fault. In other words, out of the total \(v l^2\) field elements representing a vinegar variable set, the attacker does not know the indices of the fixed elements or their values.

Let us consider the \(j\)-th linear system. To solve this system, the attacker could specialize \(l o - \Gamma + 1\) variables (i.e., assign random values to \(l o - \Gamma + 1\) variables) and then solve the resulting linear system with \(\Gamma - 1\) equations and \(\Gamma - 1\) unknowns. Consequently, the probability of finding the correct solution for this particular system is approximately \(p_j^{\Gamma - l o - 1}\). However, the attacker will not be able to verify whether a given solution is correct.

Therefore, \(\Gamma\) must be chosen such that both \(p_{\text{fail}}\) and the probability of successfully executing a key recovery attack via fault injection is negligible. Table~\ref{tab:values} provides possible values for \(\Gamma\) and \(\Lambda\) corresponding to each SNOVA parameter set.

\begin{table}[h]
\centering
\caption{Suggested values for $\Gamma$ and $\Lambda$.}
\label{tab:values}
\begin{tabular}{c|c|c|c|c}
\toprule
  Security&  & &  \\ 
level& $(v, o, q, l, \lambda)$ & $\Gamma$ & $\Lambda$   \\

\midrule
\multirow{3}{*}{I}   & $(37, 17, 16, 2, 128)$ & $10$  &  $34$     \\
                     & $(25, 8, 16, 3,128)$  & $10$  & $24$    \\
                     & $(24, 5, 16, 4,128)$  & $6$  & $20$      \\ 
\midrule
\multirow{3}{*}{III} & $(56, 25, 16, 2,192)$ & $14$ & $50$    \\
                     & $(49, 11, 16, 3, 192)$ & $9$& $33$      \\
                     & $(37, 8, 16, 4, 192)$  & $8$  & $32$     \\ 
\midrule
\multirow{3}{*}{V}   & $(75, 33, 16, 2, 256)$ & $15$ & $66$   \\
                     & $(66, 15, 16, 3, 256)$ & $14$& $45$    \\
                     & $(60, 10, 16, 4, 256)$ & $9$  & $40$      \\ 
\bottomrule
\end{tabular}

\end{table}

\section{Conclusion}
\label{sec:conclusion}
In this paper, we presented several fault attack strategies against the SNOVA cryptographic scheme. Initially, we proposed two methods of executing a fault attack, showing that our novel key recovery algorithm can recover the secret key with as few as 22 to 34 faulty signatures at the lowest security level, and up to 42 or 68 signatures at the highest level. Our experiments, implemented in both SAGE and C, demonstrated the efficiency of our algorithm under varying fault conditions.

In addition to these earlier methods, we introduced a new fault-assisted reconciliation attack in~\cref{sec:Fault_many}. This attack leverages induced transient faults to recover the secret key space by solving a quadratic system. The attack was evaluated using the lowest security parameter set for SNOVA, and the results indicated a high success rate under specific probability conditions for fault occurrence. Our experiments validated the feasibility of this attack, emphasizing its potential to weaken SNOVA's security when fault injections are possible.

To mitigate these vulnerabilities, we proposed a lightweight countermeasure that can effectively reduce the probability of successful key recovery without significantly impacting SNOVA's performance. This countermeasure is flexible and scalable, making it applicable across various SNOVA parameter sets.

Our findings underline the importance of robust fault-resistant implementations in post-quantum cryptographic schemes such as SNOVA. Future work could focus on further optimizing the countermeasures and exploring the impact of these attacks on other cryptographic systems.


\bibliographystyle{plain} 
\bibliography{refs} 

\begin{thebibliography}{10}

\bibitem{agoyan2010efficient}
Manuel Agoyan, Jean-Michel Dutertre, Philippe Hoogvorst, Emmanuel Jaulmes,
  Alfredo Tria, and Florent Valette.
\newblock Efficient fault attacks on {AES}.
\newblock In {\em IEEE Workshop on Fault Diagnosis and Tolerance in
  Cryptography}, pages 16--23. IEEE, 2010.

\bibitem{aulbach2023}
Tobias Aulbach and Florian K{\"o}nig.
\newblock Exploring the implementation security of the post-quantum signature
  scheme {MAYO}.
\newblock In {\em Workshop on Fault Diagnosis and Tolerance in Cryptography -
  FDTC 2023}, pages 10--30, 2023.

\bibitem{aulbach2022}
Tobias Aulbach, Florian K{\"o}nig, and J{\"u}rgen Kra{\"a}mer.
\newblock Recovering {Rainbow}’s secret key with a first-order fault attack.
\newblock In {\em Progress in Cryptology - AFRICACRYPT 2022}, pages 348--368,
  2022.

\bibitem{MAYO21}
Ward Beullens.
\newblock {MAYO:} practical post-quantum signatures from oil-and-vinegar maps.
\newblock In Riham AlTawy and Andreas H{\"{u}}lsing, editors, {\em Selected
  Areas in Cryptography - 28th International Conference, {SAC} 2021, Virtual
  Event, September 29 - October 1, 2021, Revised Selected Papers}, volume 13203
  of {\em Lecture Notes in Computer Science}, pages 355--376. Springer, 2021.

\bibitem{Beu24}
Ward Beullens.
\newblock Improved cryptanalysis of {SNOVA}.
\newblock Cryptology {ePrint} Archive, Paper 2024/1297, 2024.

\bibitem{mayo}
Ward Beullens, Fabio Campos, Sof\'{i}a Celi, Basil Hess, and Matthias~J.
  Kannwischer.
\newblock {MAYO}, June 2023.
\newblock Available at \url{https://pqmayo.org/assets/specs/mayo.pdf}.

\bibitem{biham1997differential}
Eli Biham and Adi Shamir.
\newblock Differential fault analysis of secret key cryptosystems.
\newblock {\em Advances in Cryptology—CRYPTO’97}, pages 513--525, 1997.

\bibitem{BUSS1999572}
Jonathan~F Buss, Gudmund~S Frandsen, and Jeffrey~O Shallit.
\newblock The computational complexity of some problems of linear algebra.
\newblock {\em Journal of Computer and System Sciences}, 58(3):572--596, 1999.

\bibitem{cryptoeprint:2024/1770}
Daniel Cabarcas, Peigen Li, Javier Verbel, and Ricardo Villanueva-Polanco.
\newblock Improved attacks for {SNOVA} by exploiting stability under a group
  action.
\newblock Cryptology {ePrint} Archive, Paper 2024/1770, 2024.

\bibitem{cryptographicestimators}
Andre Esser, Javier~A. Verbel, Floyd Zweydinger, and Emanuele Bellini.
\newblock \{CryptographicEstimators\}: a software library for cryptographic
  hardness estimation.
\newblock {\em \{IACR\} Cryptol. ePrint Arch.}, page 589, 2023.

\bibitem{furue2022}
Hiroshi Furue, Yuichi Kiyomura, and Takashi Takagi.
\newblock A new fault attack on uov multivariate signature scheme.
\newblock In {\em Post-Quantum Cryptography - PQCrypto 2022}, pages 124--143,
  2022.

\bibitem{10.1007/978-3-642-25405-5_1}
Yasufumi Hashimoto, Tsuyoshi Takagi, and Kouichi Sakurai.
\newblock General fault attacks on multivariate public key cryptosystems.
\newblock In Bo-Yin Yang, editor, {\em Post-Quantum Cryptography}, pages 1--18,
  Berlin, Heidelberg, 2011. Springer Berlin Heidelberg.

\bibitem{hashimoto2011}
Yoshinori Hashimoto, Tsuyoshi Takagi, and Kouichi Sakurai.
\newblock General fault attacks on multivariate public key cryptosystems.
\newblock {\em Post-Quantum Cryptography}, pages 1--18, 2011.

\bibitem{cryptoeprint:2024/096}
Yasuhiko Ikematsu and Rika Akiyama.
\newblock Revisiting the security analysis of {SNOVA}.
\newblock Cryptology ePrint Archive, Paper 2024/096, 2024.
\newblock \url{https://eprint.iacr.org/2024/096}.

\bibitem{jendral2024p}
Sönke Jendral and Elena Dubrova.
\newblock {MAYO} key recovery by fixing vinegar seeds.
\newblock Cryptology {ePrint} Archive, Paper 2024/1550, 2024.

\bibitem{kim2014flipping}
Yoongu Kim, Peter Daly, Jeremie Kim, Christopher Fallin, Ji~Hye Lee, Donghyuk
  Lee, Chris Lusky, Justin Meza, and Onur Mutlu.
\newblock Flipping bits in memory without accessing them: An experimental study
  of dram disturbance errors.
\newblock {\em ACM SIGARCH Computer Architecture News}, 42(3):361--372, 2014.

\bibitem{kramer2019}
Joachim Kr{\"a}mer and Maurus Loiero.
\newblock Fault attacks on uov and rainbow.
\newblock {\em Constructive Side-Channel Analysis and Secure Design}, pages
  193--214, 2019.

\bibitem{cryptoeprint:2024/110}
Peigen Li and Jintai Ding.
\newblock Cryptanalysis of the {SNOVA} signature scheme.
\newblock Cryptology ePrint Archive, Paper 2024/110, 2024.
\newblock \url{https://eprint.iacr.org/2024/110}.

\bibitem{SNOVA}
Chun-Yen~Chou Lih-Chung~Wang, Jintai Ding, Yen-Liang Kuan, Ming-Siou Li, Bo-Shu
  Tseng, Po-En Tseng, and Chia-Chun Wang.
\newblock Snova: Proposal for nistpqc: Digital signature schemes project.
\newblock Proposal for NISTPQC: Digital Signature Schemes project, 2023.
\newblock \url{https://snova.pqclab.org/}.

\bibitem{c_star}
Tsutomu Matsumoto and Hideki Imai.
\newblock Public quadratic polynomial-tuples for efficient
  signature-verification and message-encryption.
\newblock In D.~Barstow, W.~Brauer, P.~Brinch~Hansen, D.~Gries, D.~Luckham,
  C.~Moler, A.~Pnueli, G.~Seegm{\"u}ller, J.~Stoer, N.~Wirth, and Christoph~G.
  G{\"u}nther, editors, {\em Advances in Cryptology --- EUROCRYPT '88}, pages
  419--453, Berlin, Heidelberg, 1988. Springer Berlin Heidelberg.

\bibitem{mus2020}
Kaan Mus, Seungwon Shin, and Berk Sunar.
\newblock Quantumhammer: A practical hybrid attack on the luov signature
  scheme.
\newblock In {\em ACM SIGSAC Conference on Computer and Communications
  Security}, pages 1071--1084, 2020.

\bibitem{cryptoeprint:2024/1374}
Shuhei Nakamura, Yusuke Tani, and Hiroki Furue.
\newblock Lifting approach against the {SNOVA} scheme.
\newblock Cryptology {ePrint} Archive, Paper 2024/1374, 2024.

\bibitem{Patarin95}
Jacques Patarin.
\newblock Cryptanalysis of the matsumoto and imai public key scheme of
  eurocrypt'88.
\newblock In Don Coppersmith, editor, {\em Advances in Cryptology - {CRYPTO}
  '95, 15th Annual International Cryptology Conference, Santa Barbara,
  California, USA, August 27-31, 1995, Proceedings}, volume 963 of {\em Lecture
  Notes in Computer Science}, pages 248--261. Springer, 1995.

\bibitem{cryptoeprint:2014/337}
Kenneth~G. Paterson, Jacob C.~N. Schuldt, and Dale~L. Sibborn.
\newblock Related randomness attacks for public key encryption.
\newblock Cryptology ePrint Archive, Paper 2014/337, 2014.
\newblock \url{https://eprint.iacr.org/2014/337}.

\bibitem{sayari2021}
Anis Sayari, Denis Butin, and Thomas Fuhr.
\newblock Practical fault injection attacks on multivariate signature schemes.
\newblock In {\em Constructive Side-Channel Analysis and Secure Design - COSADE
  2021}, pages 101--121, 2021.

\bibitem{shim2009}
Ki~Shim and Bum~Kyu Koo.
\newblock Full key recovery on uov with fault injection.
\newblock {\em ICISC 2009}, pages 123--140, 2009.

\bibitem{snova_modified}
Lih-Chung Wang, Chun-Yen Chou, Jintai Ding, Yen-Liang Kuan, Jan~Adriaan
  Leegwater, Ming-Siou Li, Bo-Shu Tseng, Po-En Tseng, and Chia-Chun Wang.
\newblock A note on the {SNOVA} security.
\newblock Cryptology {ePrint} Archive, Paper 2024/1517, 2024.

\bibitem{cryptoeprint:2022/1742}
Lih-Chung Wang, Po-En Tseng, Yen-Liang Kuan, and Chun-Yen Chou.
\newblock A simple noncommutative uov scheme.
\newblock Cryptology ePrint Archive, Paper 2022/1742, 2022.
\newblock \url{https://eprint.iacr.org/2022/1742}.

\end{thebibliography}

\appendix
\section{Implementation details}
\label{app:code_gf}

\begin{lstlisting}[language=C, label=lst:getf16, caption=Generate random elements of $\mathbb{F}_{16}$.]
// Function to generate random field element
uint8_t random_field_element() {
    return rand() % 16; // Adjust based on field size (for GF(2^k), mod appropriately)
}

// Function to generate binomial random variable (Bernoulli trial)
uint8_t binomial_trial(double prob) {
    double r = (double)rand() / RAND_MAX;
    return r < prob ? 1 : 0;
}

// Function equivalent to get_F16
void get_F16(int v, int o, int l, uint8_t *I, uint8_t *x, double prob) {
    int size = v * l * l;
    // Generate I array with random field elements
    for (int i = 0; i < size; i++) {
        I[i] = random_field_element();
    }
    // Generate x array with binomial random variables
    for (int i = 0; i < size; i++) {
        x[i] = binomial_trial(prob);
    }
}
\end{lstlisting}

\section{Agorithm \ref{fault:esp}'s runtime complexity}

\label{runtime_algorithm_fault}

Given a homogeneous multivariate quadratic map $\mathcal{P}: \mathbb{F}^N_q \rightarrow \mathbb{F}^M_q$,  we denote $\texttt{MQ}(N,M,q)$ the field multiplications required to find a non-trivial solution $u$ satisfying $\mathcal{P}(u) = a\in \mathbb{F}^M_q$ if such solution exists. The runtime complexity of Algorithm \ref{fault:esp} is bounded by \begin{equation}
\label{compl_key_recvery_rec_fault}
    \mathcal{O}(q\sum_{\beta \in \mathcal{C}} \sum_{\gamma \in \Gamma_{\beta}} {lv \choose \gamma}\cdot \texttt{MQ}(lv-\gamma,ml^2,q) )
\end{equation} field multiplications.

\end{document}